\documentclass[reprint,amsmath,amssymb,aps,prb,floatfix,]{revtex4-2}
\usepackage{graphicx}
\usepackage{dcolumn}
\usepackage{bm}
\usepackage{siunitx}
\usepackage{hyperref}  
\usepackage{xcolor}  
\usepackage{float} 
\newcommand{\appsection}[1]{\section{\MakeUppercase{#1}}}

\newcommand{\abm}{\textcolor{black}}
\begin{document}
\preprint{Physical Review Applied, {\itshape J. Frechilla et al.}}
\title{Anisotropy by design in superconducting Nb thin films\\ via ultrashort pulse laser irradiation}%
\author{Javier Frechilla$^1$}
\author{Nicolas Lejeune$^2$}
\author{Elena Martínez$^1$}
\author{Emile Fourneau$^2$}
\author{Alejandro Frechilla$^1$}
\author{Sergio Martín$^1$}
\author{Leonardo R Cadorim$^{3,4}$}
\author{Luis A  Angurel$^1$}
\author{Germán F de la Fuente$^1$}
\author{Alejandro V Silhanek$^2$}
\author{Milorad V Milo\v{s}evi\'{c}$^4$}
\author{Antonio Badía-Majós$^{1,}$}
\email{anabadia@unizar.es}
\affiliation{{$^1$Instituto de Nanociencia y Materiales de Aragón (INMA), CSIC-Universidad de Zaragoza}, E-50018, Spain}
\affiliation{{$^2$ Département de Physique, Q-MAT, CESAM, Université de Liège}, B-4000, Belgium}
\affiliation{{$^3$Departamento de Física, Universidade Estadual Paulista}, Bauru/SP, 17033-360, Brazil}
\affiliation{{$^4$Department of Physics, University of Antwerp, Groenenborgerlaan 171, B-2020, Belgium}}
\date{\today}
\begin{abstract}
\abm{The ability to fabricate anisotropic superconducting layers {\itshape {\`a} la carte} is desired in technologies such as fluxon screening or removal in field-resilient devices, flux lensing in ultra-sensitive sensors, or in templates for imprinting magnetic structures in hybrid magnetic/superconducting multilayers. In this work, we demonstrate tailored superconductivity} in polycrystalline niobium thin films exposed to femtosecond ultraviolet laser pulses. The samples exhibit significant changes in their superconducting properties, directly connected with the observed topography, crystallite geometry, and lattice parameter modifications. On the mesoscopic scale, quasi-parallel periodic ripple structures (about 260 nm of spatial period) gradually form on the film surface by progressively increasing the laser energy per pulse, $E_{\rm p}$. This gives way to a stepwise increase of the critical current anisotropy and magnetic flux channeling effects along the ripples. As demonstrated in our resistive and inductive measurements, these superstructures determine the electromagnetic response of the sample within the regime dominated by flux-pinning. Time-dependent Ginzburg-Landau simulations corroborate the topographical origin of the customized anisotropy. Concurrently, intrinsic superconducting parameters (critical field and temperature) are moderately and isotropically depressed upon increasing $E_{\rm p}$, as is the lattice parameter of Nb. These findings \abm{promote pulsed laser processing as a flexible, one-step, and scalable {lithography-free} technique for versatile surface functionalization in microelectronic superconducting technology.}
\end{abstract}
\maketitle
%
%
\section{\label{sec:intro}Introduction}

Niobium thin films are key components of a wide range of technologies, such as those involved in Josephson junctions, superconducting quantum interference devices (SQUID), microwave resonators, or radiofrequency superconducting cavities. The optimization and control of the crystallographic and superconducting properties of the films is thus essential and has been well reported in the literature~\cite{bose_05,gubin_05,choi_20,imamura_92,liu_09,gao_22,zhong_23, torres_24, Karuppannan_23, Liu_2025}. Specifically, it has been found that the intrinsic superconducting properties of Nb (i.e., superconducting energy gap, magnetic penetration depth, and critical temperature, $T_{\rm c}$) are dependent on dimensional parameters~\cite{bose_05,gubin_05,choi_20}. Thus, a noticeable reduction of $T_{\rm c}$ was observed when grain size~\cite{bose_05} or film thickness~\cite{gubin_05} reduces below 30-50 nm. This behavior was attributed to finite-size effects (that is, length scales close to the coherence length of the material $\xi$), to changes in the electronic density of states or to the increase of film disorder. In general, induced stress and structural disorder, which vary with growth conditions of the film, may affect the critical temperature, the residual resistance ratio (RRR), and/or the lattice parameter of the Nb layer~\cite{bose_05,gubin_05,imamura_92,liu_09,gao_22,zhong_23,Liu_2025}. 

Along with the variation of the intrinsic superconducting properties, nanopatterning of Nb films, often required in device fabrication processes, may also produce changes in their magnetic flux pinning landscape, often relevant for applications. In particular, engineering of dedicated sample architectures at the nanometer scale opens the possibility of manipulating magnetic flux and has enabled the appearance of the so-called {\itshape fluxonic devices}~\cite{savelev_02,lee_99,villegas_03,silhanek_03,silhanek_10,cuppens_11,dobrovolski_15,vlasko-vlasov_16}, also emerging as a promising technology in cutting-edge applications as the quantum computation systems. 
\abm{Thus, the manipulation of magnetic flux by dedicated intervention in the mesoscopic scale, in particular through anisotropic surface patterns, allows to fabricate such devices as rectifiers~\cite{villegas_03}, filters\cite{dobrovolski_15}, and fluxon lenses and diodes~\cite{zhu_04}.}
Another topical issue in superconducting devices is the unwanted presence of penetrating Abrikosov vortices, which may cause noise, distort the response of Josephson junctions, \abm{or even take place in the form of catastrophic flux avalanches in radiofrequency devices~\cite{lejeune_23}. Among other strategies to drive flux out of the device, the proposal of using patterned superconducting layers with ratchet thickness profiles (typical length scales in the submicron range) came into the scene in the late 90's~\cite{lee_99}, and is still under development~\cite{koshelev_25} from the theoretical point of view, although it has not yet received a satisfactory experimental realization. Other authors have proposed to create field-resilient superconducting devices by engraving protective \abm{flux moats}~\cite{bermon_83,song_09,pan_25} which could also be fabricated by means of anisotropic frames surrounding the protected area.}
\abm{Finally, related to the extensively investigated interaction between superconducting vortices and other topological objects such as magnetic skyrmions~\cite{reichardt_22}, one may find proposals of using fluxon lattices as templates for imprinting skyrmions in neighbouring layers~\cite{delvalle_15}, as well as controlling the motion of such skyrmions~\cite{Menezes2019}. Again, the control of magnetic flux pathways through anisotropic surface structures in the superconductor could act as a mask to control the neighboring skyrmion dynamics.}

An emerging technique to produce nanopatterns and other surface modifications is based on the irradiation with femtosecond (fs) pulse lasers. This method, of growing interest in recent years, has already been applied to achieve performance improvements of very diverse thin films by inducing chemical, structural, and/or physical modifications~\cite{BONSE_23,  chen_23, sharif_22, lapidas_25, Skoulas_21, Vilkevicius_24}.
Related to this, in previous works \cite{cubero_20,cubero_nano_20}, it was shown that surface nanostructures can modify the superconducting properties of Nb sheets, and we conceived the functionalization of Nb thin films by fabricating artificial anisotropic surface topographies devised as rails for the magnetic flux \cite{badia_24}. We showed that surface nanopatterning by ultra-short pulsed-laser irradiation may result in anisotropic electromagnetic behavior for superconducting Nb films. More specifically, by selecting adequate conditions, laser-induced periodic surface structures (LIPSS), also named ripples, can be generated on the sample's surface. The current carrying capacity of Nb films is enhanced along the direction of thereby induced quasi-parallel ripples. As a matter of fact, by properly assembling conjoined domains of ripple structures, we could demonstrate the possibility of building flux pathways in the films~\cite{martinez_25}.
Nevertheless, together with this appealing property, it was noticed that the intrinsic nature of the underlying superconducting material may be undermined. In particular, a faint reduction of $T_{\rm c}$ was reported~\cite{martinez_25}, suggesting a correlation with the microstructural modifications induced by laser irradiation. Thus, in samples with a non-homogeneous laser-patterning motif, magneto-optical imaging (MOI) experiments revealed a locally changing value of $T_{\rm c}$ for zones with different microscopic order.

In this \abm{work}, we present an oriented investigation of phenomenology related to the formation of LIPSS \abm{in Nb thin films}. 
%
\abm{Special emphasis is placed on the ability to control the surface modifications induced by the laser pulses, as well as the implications for the performance of the superconductor.}
Furthermore, focusing on the relevance of patterning in the mesoscopic scale (motif size $\Lambda$ is comparable to the typical lengths of superconductivity $\xi,\lambda$ for Nb), we have performed time-dependent Ginzburg-Landau (TDGL) simulations that support our interpretation of the underlying physics in the samples with surface undulations.

The article is organized as follows. In Sec.~\ref{sec:methods}, we give details on the sample fabrication and laser patterning methods, as well as about the experimental characterization techniques (SEM, TEM, XRD, SQUID magnetometry, and MOI). Sec.~\ref{sec:theory} puts forward the details of the TDGL simulation method. Sec.~\ref{sec:results} presents the main findings, showing the most relevant microstructural (lattice parameters, surface topography) and physical properties (critical fields, currents, and temperatures) of the irradiated films as compared to the pristine samples. Finally (Sec.~\ref{sec:conclusions}), in view of the observed phenomenology, we extract some findings on the influence of laser-processing techniques on the superconducting properties.

\section{\label{sec:methods} Experimental details} 

\subsection{\label{sec:film_growth_laser_methods}Sample fabrication and laser nanopatterning}

{Nb thin films ($d \approx 190$~\unit{\nano\metre} of thickness)} were deposited onto Si/SiO$_2$ wafers, using radio-frequency magnetron sputtering. The Nb physical vapor deposition was conducted at a rate of {0.1 nm/s} under an Ar pressure of ${5\; {\rm mTorr}}$, after pumping down the chamber to a base pressure of $4\cdot 10^{-8}$ Torr.

\begin{figure}[h]
\includegraphics[width=0.45\textwidth]{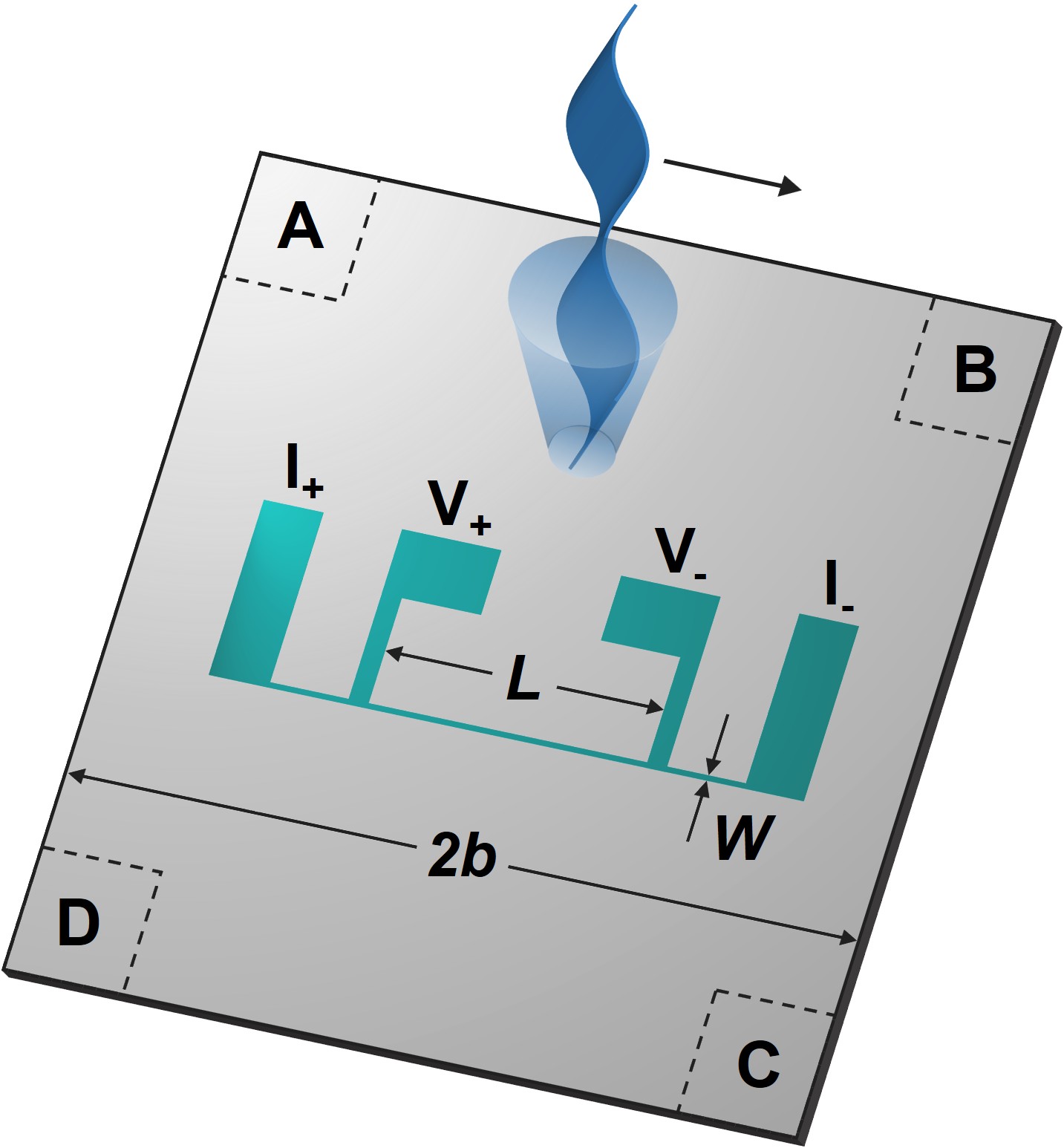}
\caption{\label{fig:figure_sketch}A) {Sketch of the sample geometries and laser scanning process. Square films $2b\times 2b$ were laser cut. On some of them, the Nb layer was selectively removed by laser ablation (pulse energy $E_{\rm p}=6$~\unit{\micro\joule}), defining a transport bridge geometry. Subsequently, the surface was corrugated by laser irradiation ($E_{\rm p}=3.4$~\unit{\micro\joule}). Ultrasonic wire bonding was performed on the four corners (A--D) for the square films or on top of the specific pads ($V_{\pm}, I_{\pm}$) of the bridges for resistive measurements. For visual purposes, the laser spot is not to scale.}}
\end{figure}

A pulsed laser (Light Conversion, Vilnius, Lithuania, model Carbide CB3-40W) was used for cutting, ablating, and nanopatterning the samples, as described below: 

{i) {\itshape Sample cutting}: To begin with, square samples were laser-cut from the wafers using wavelength {$\lambda =515\, {\rm nm}$}, pulse width {$\tau_p={249}\, {\rm fs}$}, pulse repetition frequency ${f_{\rm rep}=10\, {\rm kHz}}$ and a laser scanning speed ${v=5\, {\rm mm/s}}$. After the machining process, the samples were ultrasound cleaned in isopropanol for {15~min}. 

Two types of samples were prepared in this study, which will be named after ``film" and ``bridge" (see Fig.~\ref{fig:figure_sketch} and Table~\ref{tab:table_1}). The former are square-shaped samples, whereas the latter are dedicated geometries for resistive measurements.

{ii) {\itshape Fabrication of bridges}: As sketched in Fig.~\ref{fig:figure_sketch}, the bridges have reduced widths, $W$, of a few tens of microns and were fabricated from the square samples by complete removal of the Nb layer on selected areas using the following conditions for laser ablation: {{$\lambda =343\, {\rm nm}$}, {$\tau_p={238}\, {\rm fs}$},  ${f_{\rm rep}=20\, {\rm kHz}}$, ${v=125\, {\rm mm/s}}$, distance between lines {$\delta_l=6\,$\textmu{m}}} and pulse energy $E_{\rm p} \approx 6.0 \,$\textmu{J}.

As it will be later discussed, the bridge geometry conveys several advantages: (a) a lower level of transport current is needed, (b) the analysis of experiments is simplified (the orientation of the current density vector {\bf J} is known {\itshape a priori}) and (c) one gets rid of extrinsic spurious effects, concomitant of anisotropic 2D configurations (see Appendix~\ref{app:A3}). 

\begin{table*}
\caption{\label{tab:table_1}Description of the two types of samples used in this study (square film and bridge samples) and laser pulse energies used for surface nanopatterning. The other (shared) laser irradiation parameters are given in the text.}

\begin{ruledtabular}
\begin{tabular}{ccccc}

Sample name&Description &Pulse energy& $E_{\text{p}}/E_{\text{LIPSS}}$ \\
{}&{} &$E_{\text{p}}$(\unit{\micro\joule})&
\\ \hline
FS0 &Square, pristine  &  0 & 0  \\
FSj &Square samples ($j = 1, 2, 3,\dots$) & $1.2-3.8$ & $0.35-1.10$ \\
FSL &Square, fully covered by LIPSS  & 3.4 & 1 \\
BS0 &Bridge, pristine  &  0 & 0  \\
BS$_{\rm PAR}$ &Bridge, parallel LIPSS &  3.4 & 1 \\
BS$_{\rm PERP}$ &Bridge, perpendicular LIPSS  &  3.4 & 1 \\
\end{tabular}
\end{ruledtabular}
\end{table*}}

{iii) {\itshape Surface nanopatterning}: Finally, as also sketched in Fig.~\ref{fig:figure_sketch}, the surface of film and bridge samples was nanopatterned to produce LIPSS by laser scanning the sample using the following parameters: {$\lambda =343\, {\rm nm}$}, {$\tau_p={238}\, {\rm fs}$}, ${f_{\rm rep}=20\, {\rm kHz}}$, ${v=125\, {\rm mm/s}}$, {$\delta_l=8\,$\textmu{m}} and pulse energy, $E_{\rm p}=E_{\rm LIPSS} \approx 3.4 \,$\textmu{J}.  {For the selected wavelength and at the established working distance, our laser beam exhibited a spatially Gaussian energy profile with an elliptical spot of semi-axes {$18 \times 30\,$\textmu{m}$^2$} for the $1/{\rm e}^2$ decay distance.} This set of laser-irradiation parameters, critical for the generation of uniform LIPSS structures, was refined on a previous study~\cite{martinez_25}. 

Two bridge samples were irradiated to generate LIPSS either parallel or perpendicular to the path between voltage pads (i.e., parallel or perpendicular to the current applied in transport measurements). In addition, we irradiated a series of square samples (of width $2b \approx 3 {\rm ~or~} 4$~\unit{\milli\metre}) with varying $E_{\rm p}$ values ranging from 0.35 to 1.1$E_{\rm LIPSS}$ (the rest of laser conditions remaining unchanged). In all cases, a controlled Ar atmosphere (gas flow in an open chamber) was used, not only for reducing possible oxidation induced by laser-processing in air, but also as a means of removing debris particles through the gas flow.

\subsection{\label{sec:microstructural_method}Microstructural observations}

\subsubsection{\label{sec:SEM_TEM}Electron microscopy}

Surface characterization was carried out using a MERLIN field-emission scanning electron microscope (FE-SEM) (Carl Zeiss in Jena, Germany), equipped with an energy dispersive X-Ray spectroscopy (EDS) system from Oxford Instruments (Abingdon, UK). The FE-SEM was operated at 5 kV and utilized secondary electron (SE), in-lens, and energy-selective backscattered (ESB) detectors.  Cross-sectional TEM images were taken with a Tecnai F30 microscope of FEI (Lincoln, NE, USA). With this aim, several lamellas were prepared using a Focused Ion Beam (FIB) in a Dual Beam Helios 650 apparatus of FEI. Prior to the FIB-lamella extraction, samples were coated with C-Pt protective layers.

\subsubsection{\label{sec:XRD}X-ray diffraction}

X-ray diffraction (XRD) analyses were conducted on a series of thin films irradiated with different pulse laser energies in order to assess the influence of the radiation on the material’s crystal structure. Measurements were performed using a PANalytical Empyrean system configured in Bragg-Brentano geometry, equipped with a copper anode X-ray source ($\lambda$ = 1.5418 $\text{\AA}$) and a PIXcel linear detector covering an angular range of 3.3$^\circ$. To collect the complete signal from the film within the 30–45$^\circ$ detector angle range while avoiding the strong reflection from the single-crystal silicon substrate at 2$\theta$ = 66.2$^\circ$, the incidence angle ($\omega$) was deliberately tilted by 5$^\circ$ away from the symmetric position. Data processing included curve smoothing by removing the background and averaging the measured values over every 10 data points to minimize errors in the resulting curves.

The lattice parameter $a$ of the body-centered cubic Nb~\cite{XRD} was calculated for each sample from the diffraction measurements. As customary, we used Bragg's law to determine the interplanar spacing $d_{hkl}={n \cdot \lambda}/{2 \sin \theta}$, i.e.: $a = d_{hkl} \cdot \sqrt{h^2 + k^2 + l^2}$
with $n = 1$ (first-order diffraction), $\lambda$ the X-ray wavelength, $\theta$ the diffraction angle; and $h$, $k$, $l$ the Miller indices of the plane, in this case (110).

\subsection{\label{sec:EM_methods}Electromagnetic measurements}

Electric transport and inductive measurements were performed in our thin films, focused on the influence of the laser-induced modifications on their physical properties.

\subsubsection{\label{sec:elec_methods}Electrical resistivity}

The electrical resistivity of bridges and square films was measured from 3~\unit{\kelvin} to room temperature in a Quantum Design PPMS instrument (San Diego, CA, USA) with DC field (0 to 1~\unit{\tesla}) applied perpendicular to the film's surface. For the case of square films, a standard van der Pauw configuration was used. Taking advantage of the multichannel options, the experimental data were obtained simultaneously for three different samples subject to comparison. In particular, we measured $\rho (H)$ for different sets of temperatures as well as $\rho(T)$ for sets of values of the applied magnetic field. In between each set of measurements, the samples were heated above the superconducting critical temperature and allowed to cool down in a zero magnetic field.} {The $T_{\rm c}$} values of the samples were obtained from $\rho(T)$ at zero DC field, while the higher critical fields were derived from the transitions in terms of the applied field at different temperatures $\rho(H,T_{_0})$.

\subsubsection{\label{sec:squid_methods}SQUID magnetometry}

The complex AC susceptibility, $\chi_{\rm ac}(T)$, with in-phase $\chi '$ and out-of-phase $\chi ''$ components, was measured at zero DC magnetic field and AC drive magnetic field (sine wave amplitude ${\mu_{_0}h_{_0} = 10}$~\textmu T, frequency ${f = 10~{\rm hertz}}$) in a SQUID magnetometer (MPMS3 from Quantum Design, San Diego, CA, USA).   The $T_{\rm c}$ values of the square film samples were obtained from these measurements using the onset of diamagnetism criterion. The same system was used to perform hysteresis loops $M(H)$ at constant temperature using a scan length of 3 cm. The magnetic field was ramped at 2 mT/s to each set value with the non-overshoot mode. After field stabilization, the system was paused for 2 s before the measurement started.

All magnetic measurements were performed on the above-described square samples with the magnetic field applied perpendicular to the surface and after zero field cooling. For each measurement, a reset of the superconducting magnet was performed previous to sample cooling in order to minimize its remnant magnetic field.

\subsubsection{\label{sec:moi_methods}Magneto-Optical Imaging (MOI)}

The analysis of the magnetic flux penetration in the laser-patterned samples was performed using the magneto-optical Faraday imaging technique. The employed experimental configuration mimics a commercial polarization microscope. A light beam emitted by an LED lamp passed through a green filter (550 nm) and a linear polarizer before reaching a Faraday active indicator. The indicator in this study consisted of a {$3.5\,$\textmu{m}}-thick layer of Bi:YIG, epitaxially grown on a {$450\,$\textmu{m}}-thick Gd$_3$Ga$_5$O$_{12}$ (GGG) transparent substrate, with a 100 nm-thick Al mirror deposited on the optically active layer to ensure adequate reflection of the incident light.
The linearly polarized light traversed the GGG substrate and the Bi:YIG layer within the indicator, undergoing a polarization direction rotation corresponding to the local magnetic field. Subsequently, the light was reflected by the mirror, retraced its path through the indicator and the objective, and finally reached an analyzer. The collected light was then directed to a high-resolution CCD camera situated atop the microscope unit, producing an image in which the intensity contrast directly represented the magnetic field distribution. More technical details may be found elsewhere~\cite{shaw_18}.

\begin{figure*}[!]
\includegraphics[width=.9\textwidth]{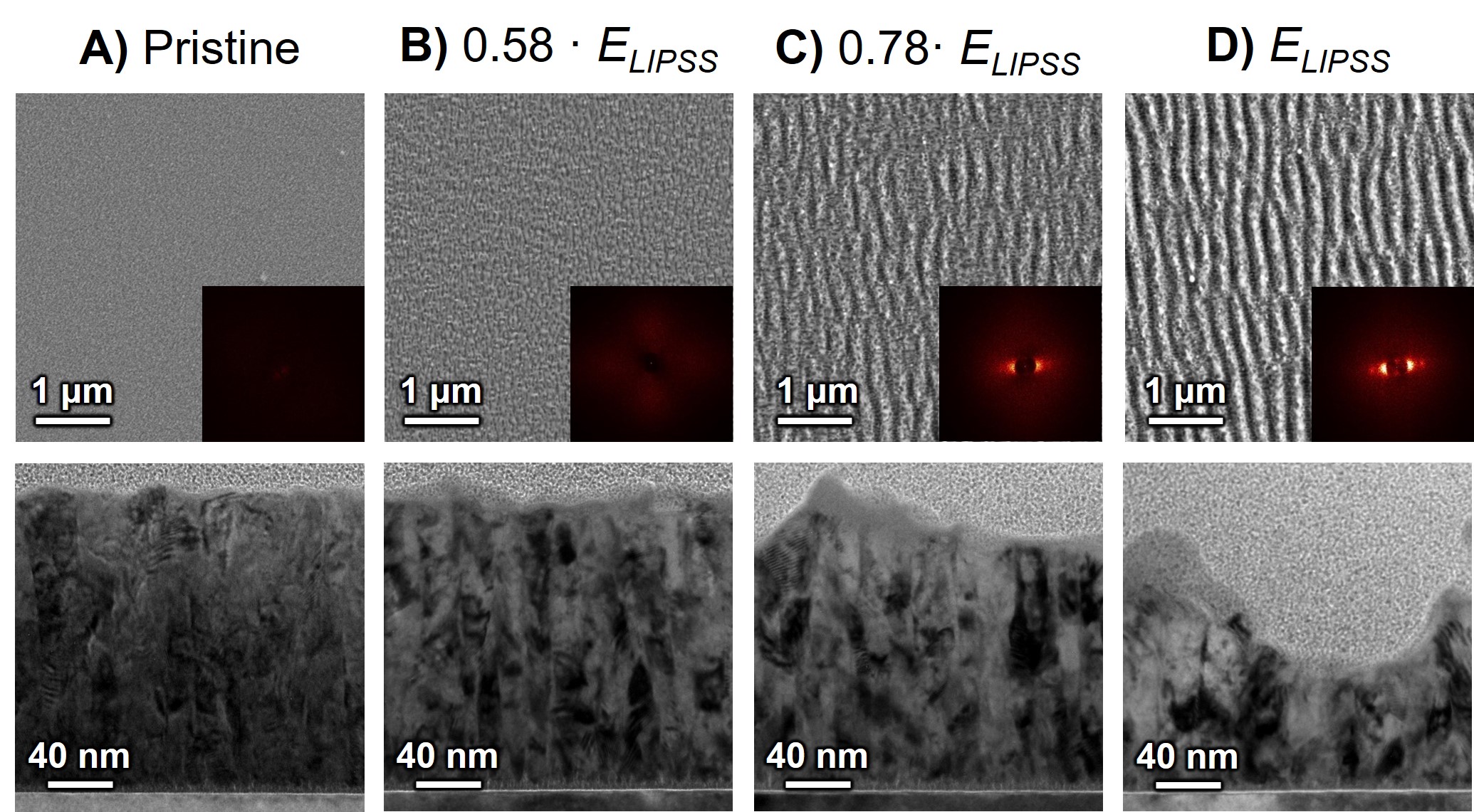}
\caption{\label{fig:figure_sem_gradual} Microstructural details of square-shaped samples that show the gradual effect of increasing the laser pulse energy. A) Pristine sample (FS0); B) and C) two samples of the series FSj irradiated with two intermediate $E_{\rm p}$ values as indicated; and D) Sample FSL whose surface is fully covered by LIPSS. Upper panels:  Top-view SEM (SE) images. Insets show 2D-FFT of SEM images ($\approx 70~\unit{\micro\meter}^2$ of area). Lower panels: Cross-sectional brightfield TEM images.} 
\end{figure*}
\section{\label{sec:theory} Modeling: TDGL theory}

To theoretically corroborate the interpretation of our experimental results, we resorted to time-dependent Ginzburg-Landau simulations. This allows us to obtain the vortex dynamics in samples with arbitrary motifs at the mesoscopic scale, i.e.: for length scales of the order of the superconducting coherence length.
The simulations were carried out through the numerical solution of the two-dimensional TDGL equation adapted to superconducting films with variable thickness \cite{chapman1996}
\begin{eqnarray}
    u\left ( \frac{\partial }{\partial t}
    +\text{i}\varphi \right ) \psi &=&  
    \frac{1}{d}\left(\mbox{\boldmath $\nabla$}-\text{i}\textbf{A}\right)\cdot d\left(\mbox{\boldmath $\nabla$}-\text{i}\textbf{A}\right)\psi \nonumber \\
    &+&\psi(1-|\psi|^2)\, ,
    \label{eq:eq1}
\end{eqnarray}
where $d \equiv d(x,y)$ is the thickness profile of the film and $u$ is a constant ($\approx 5.79$) representing the ratio of relaxation times for the amplitude and phase of the order parameter in
dirty superconductors. The order parameter $\psi$ is in units of the field-free order parameter $\psi_\infty$ at the given temperature $T$; lengths in units of the superconducting coherence length $\xi(T)$; time in units of $t_{GL}=\pi\hbar/8uk_BT_{\rm c}$; the vector potential $\bf{A}$ in units of $H_{c2}\xi$, with $H_{c2}$ being the upper critical field, and the electrostatic potential $\varphi$ in units of $\hbar/2et_{GL}$. $\varphi$ is obtained from the Poisson equation:
\begin{equation}
    \bm{\nabla}\cdot(d\bm{\nabla}\varphi) = \bm{\nabla}\cdot(d\bm{J}_s),
\end{equation}
where the supercurrent is given by:
\begin{equation}
    \bm{J}_s = \textrm{Im}\left [\bar{\psi}{(\mbox{\boldmath $\nabla$}-\text{i}\textbf{A})}\psi\right].
\end{equation}
Boundary conditions for $\psi$ are set to guarantee no supercurrent flows out of the film. The applied current is introduced through the boundary condition for $\varphi$, with $\bm{\nabla}\varphi = \bm{J}_{applied}$ along the surfaces perpendicular to the external current, while $\bm{\nabla}\varphi = 0$ is set at the remaining surfaces. As in experiments, when considered, the external magnetic field will be assumed to be applied along the $z-$axis, i.e., normal to the film. Details on the numerical discretisation can be found in Ref.~\cite{milovsevic2010}. Aiming at disclosing the role of surface undulations on the sample's physical properties, simulations were performed by neglecting flux pinning effects other than those linked to thickness variations or edge barriers.

\section{\label{sec:results}Results and discussion}

\subsection{\label{sec:patterning}Laser patterning: evolution of the microstructure}

The gradual effect on the topography and microstructure of the Nb films upon increasing the laser pulse energy may be recognized in Fig.~\ref{fig:figure_sem_gradual}. Starting from the pristine sample FS0, we first observed a slight increase in surface roughness for the sample irradiated with ${E_{\rm p}/E_{\rm LIPSS } = 0.58}$. Upon increasing this factor to about 0.7, LIPSS start nucleating non-homogeneously at first (see for example figure Fig.~\ref{fig:figure_sem_gradual}C, corresponding to ${E_{\rm p}/E_{\rm LIPSS } = 0.78}$), and progressively reach homogeneity as shown in Fig.~\ref{fig:figure_sem_gradual}D ($E_{\rm p}=E_{\rm LIPSS}$), where complete LIPSS coverage is seen. Upon further increasing $E_{\rm p}$ for about 10 $\%$, some small holes appear in the Nb films. 
\begin{figure*}[!]
\includegraphics[width=.9\textwidth]{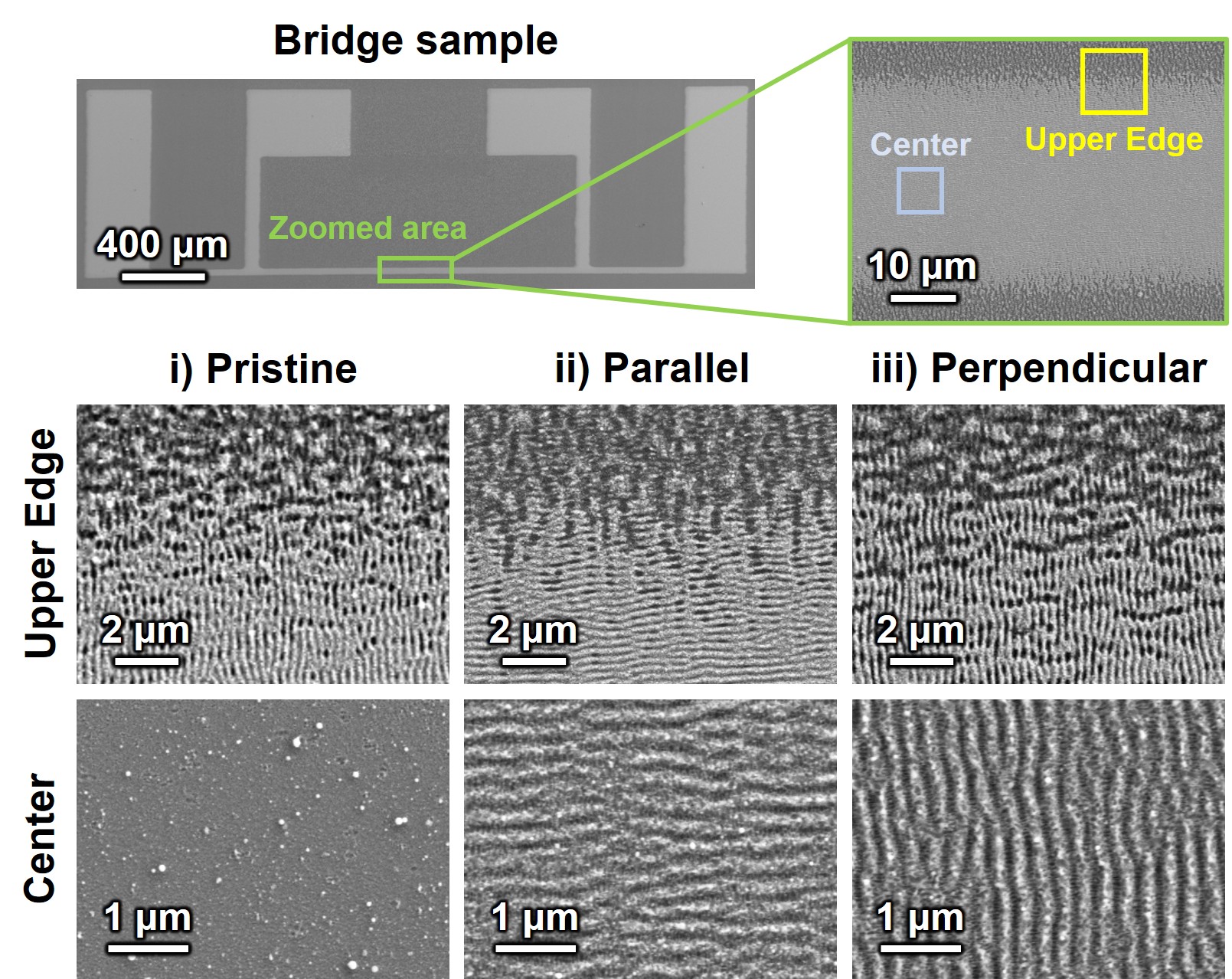}
\caption{\label{fig:figure_sem_bridges} SEM (SE) images of the surface of bridge samples: the upper panels correspond to the sample BS$_{\rm PAR}$ and show the complete circuit and a detail of the bridge between the voltage contacts. Middle and lower panels show details at higher magnification of the bridges near the center and at the edge for the three analyzed samples: BS0 (pristine), BS$_{\rm PAR}$ (parallel), and BS$_{\rm PERP}$ (perpendicular). {The unwanted laser-patterning induced LIPSS of the pristine sample (BS0) occupy a lateral strip of 12.5~$\pm$~2.5~$\mu$m (see text)}}
\end{figure*}

The 2D Fast Fourier Transform (2D-FFT) analysis shows that once formed, ripples have a similar spatial period  $\Lambda$ = 260 $\pm$ 5 nm and are aligned perpendicular to the laser polarization direction. Note that the sample represented in Fig.~\ref{fig:figure_sem_gradual}B already exhibits slight preferential (bidirectional) organization as revealed by the FFT, unlike the pristine sample. TEM images of the cross-sections of the same samples also display some gradual changes, such as a specific decrease in the film thickness, the gradual formation of the undulated structure, as well as an increase in the grain size, especially at the top of the LIPSS, upon increasing pulse energy values.

SEM images of the surfaces of the bridge samples are displayed in Fig.~\ref{fig:figure_sem_bridges}. The images were taken in the area between voltage pads (green rectangle in this figure). At the center of the pristine bridge BS0, some small particles are visible on the surface, probably generated during the machining process. Note also that unsought LIPSS are generated during the Nb layer removal around the path in this sample, as related to the laser beam lateral tails (gaussian energy distribution of the laser spot~\cite{badia_24}), having an approximate width of {12.5~$\pm$~2.5~$\mu$m}. At the center of the paths of the BS$_{\rm PAR}$ and BS$_{\rm PERP}$ samples, LIPSS are similar to those of the FSL sample but with two different orientations, eventually parallel or perpendicular to the applied current, which will flow along the horizontal direction in these images. In both samples and for the same reason as above, there is an area around the paths' edges where the LIPSS are more pronounced, producing many holes in the Nb layer. Because the local current there (in a strip of width of about 4-8 $\mu$m) will be negligible, these zones were not taken into account to estimate the path width, ${W\approx}~$ 53, 25 and 37 $\mu$m for BS0, BS$_{\rm PAR}$ and BS$_{\rm PERP}$, respectively, with estimated error of ${\approx}$ $\pm$2 $\mu$m.  

XRD experiments were conducted for a set of square film samples irradiated with different laser pulse energies in order to analyze the effects on the film's crystallographic properties.  Fig.~\ref{fig:figure_xrd_tc_gradual}A displays a gradual shift of the Nb (110) diffraction peak angle upon increasing pulse energy values. Thus, for the non-irradiated samples, this peak is at 2$\theta$ = 38.5$^{\circ}$, which corresponds approximately to the value for bulk cubic (Im3m) Nb~\cite{XRD}, and shifts to ${\approx}$ 39.0$^{\circ}$ for $E_{\rm p}=E_{\rm LIPSS }$ (FSL sample). Correspondingly, the estimated lattice parameter for the pristine sample FS0 is 3.3055 $\pm$ 0.001 \r{A}, and decreases upon increasing the pulse energy increases, first smoothly for $E_{\rm p}/E_{\rm LIPSS }< $ 0.6, and then more sharply from $E_{\rm p}/E_{\rm LIPSS }{\approx}$ 0.7, coincident with the initial generation of LIPSS on the film surface (see Fig.~\ref{fig:figure_xrd_tc_gradual}B). It is noted that the lattice shrinkage for the sample fully covered by LIPSS (FSL, $E_{\rm p}/E_{\rm LIPSS}=1$) is thus {$\approx 1.05 \%$} as compared to the pristine sample. Finally, note that there is a noticeable decrease of the diffraction peak height intensity for the sample irradiated with {$E_{\rm p}/E_{\rm LIPSS }= 1.1$}. It occurs in parallel with the appearance of holes in the Nb film for this sample. 

\begin{figure*}[!]
\includegraphics[width=.9\textwidth]{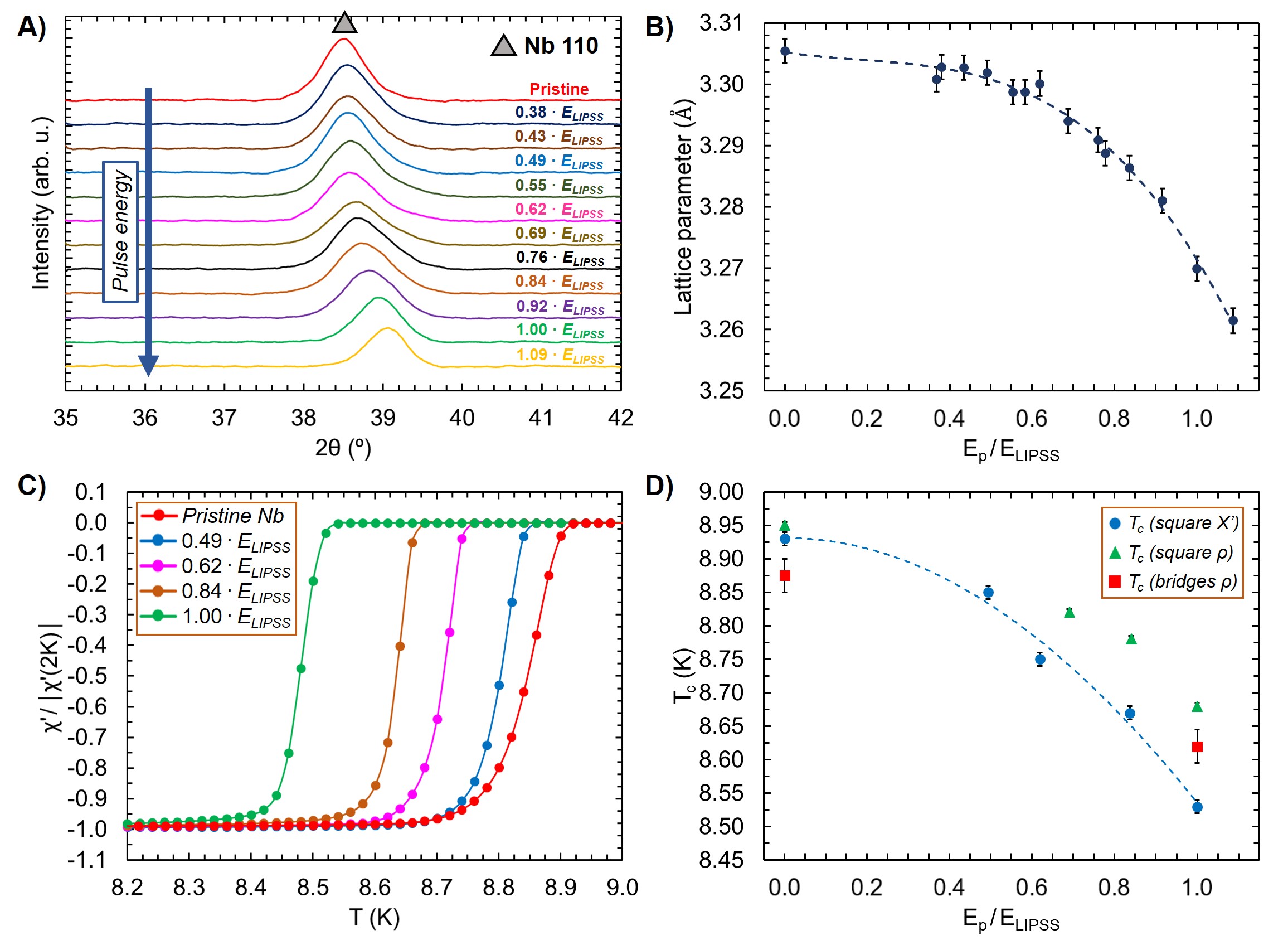}
\caption{\label{fig:figure_xrd_tc_gradual} Effect of changing the laser pulse energies: A) (110) diffraction peak of Nb. B) Estimated lattice parameter $a$ from the XRD experiments. C) $\chi'(T)$ scaled by $|\chi '$(2K)$|$, showing the superconducting-to-normal transition at zero DC field of square films processed with different pulse energies. D) $T_{\rm c}$ values estimated from susceptibility or resistivity measurements of several bridge and square samples, as indicated.  Reference values for {bulk Nb} (not displayed): {$a=3.3066(1)$~\textup{\AA}~\cite{XRD}}, $T_{\rm c}=9.25\pm 0.01$~\unit{\kelvin}~\cite{finnemore_66}. Dashed lines are guides for the eye.}
\end{figure*}

\subsection{\label{sec:resistive}{Intrinsic superconducting parameters}}
\subsubsection{\label{sec:Tc}Critical temperature}

The critical temperatures of the analyzed samples were obtained from AC susceptibility and resistivity measurements under zero DC magnetic field. Fig.~\ref{fig:figure_xrd_tc_gradual}C shows the superconducting-to-normal transition observed in $\chi'(T)$ of selected square film samples. This transition shifts towards lower temperatures upon increasing $E_{\rm p}$, but its width, $\Delta T $, defined here as the temperature interval where $\chi '$ changes between 0.05 to 0.95 of the signal measured at 2 K, does not vary significantly among the samples ($\approx$ 0.11$-$0.18 K). In the figure, $\chi'(T)$ is scaled by $|\chi '$(2K)$|$, which is close to the value expected for perfect diamagnetism in perpendicular fields, $\chi '_{\rm ideal}$ = $0.9~b/d$ ~\cite{brandt-prl_95}. $T_{\rm c}$ estimated from these measurements decreases from 8.93 K in our pristine films down to 8.53 K for $E_{\rm p}$ = $E_{\rm LIPSS}$, as displayed in Fig.~\ref{fig:figure_xrd_tc_gradual}D. The same trend is also found for the $T_{\rm c}$ values obtained from resistivity measurements in square and bridge samples, also represented in the figure for comparison purposes.} {In this case, the transition widths were {$\approx$ 0.02-0.05~\unit{\kelvin}}. The differences among datasets are likely due to the different conditions applied in each experiment: the used applied AC field amplitude ($\mu_{\rm 0}h _{\rm 0}=0.01~\mu$T for the square samples) and transport currents (1~\unit{\micro\ampere} for bridge samples and 100~\unit{\micro\ampere} for square samples), as well as to the fact that measurements are done in different instruments. Nonetheless, some general trends may be extracted. Note, for instance, that both bridge samples with LIPSS have the same critical temperature. Note also the similarity between the functions depicted in Fig.~\ref{fig:figure_xrd_tc_gradual}B and Fig.~\ref{fig:figure_xrd_tc_gradual}D, suggesting a correlation between both variables, $a$ and $T_{\rm c}$. This is in agreement with previous results~\cite{choi_20} in which decreasing trends of $T_{\rm c}$ were observed in Nb films with increasing compressive and tensile strains (with stronger dependence for the latter). In brief, one may conclude that approaching $E_{LIPSS}$, the crystal lattice parameter decreases. This implies a compressive stress which in turn leads to a decrease of $T_{\rm c}$.}

\subsubsection{\label{sec:ht}Reversible magnetization: H--T line}

\begin{figure}[!]
\includegraphics[width=.45\textwidth]{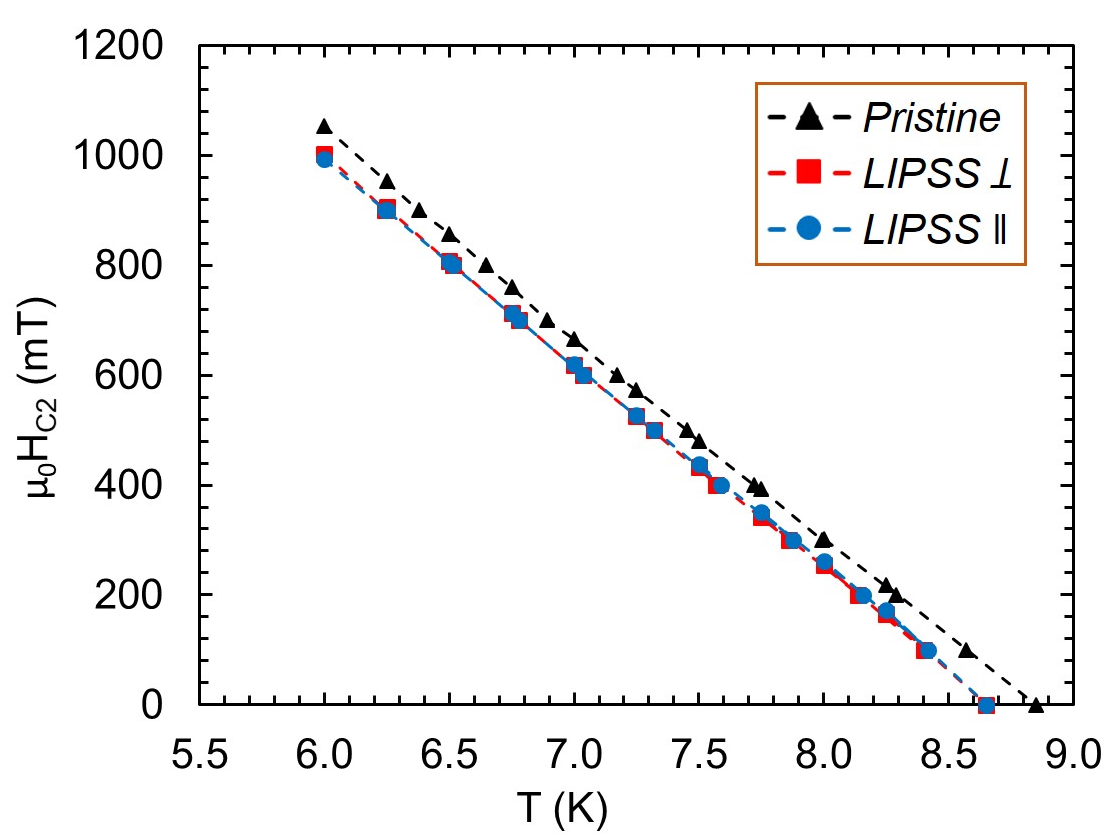}
\caption{\label{fig:figure_hc_tc}  Fraction of the equilibrium phase diagram (upper critical field $H_{c2}$) obtained from resistive measurements. The zero field values, i.e. $T_{\rm c}$, are those displayed in Fig.~\ref{fig:figure_xrd_tc_gradual} and were obtained with $I=1$~\unit{\micro\ampere}).
} 
\end{figure}

Below, we describe the effects of laser irradiation on the reversible superconducting properties of our Nb films. The analysis is based on the magnetoresistive behavior $R(H,T)$ for the samples BS$_{\rm PAR}$ and BS$_{\rm PERP}$ and pristine for low transport currents. Primary experimental data are provided in Appendix~\ref{app:appA}, with the derived physical parameters presented in Fig.~\ref{fig:figure_hc_tc}. As shown in the direct experimental data, we have investigated the transition to the normal state induced by field and temperature. 

The main result is that the equilibrium properties of our laser-treated samples are basically independent of the current density orientation relative to the LIPSS, while always  (slightly) depressed as compared to the pristine sample, i.e.: the resistance curves for the pristine sample are always shifted to higher fields and temperatures. In order to quantify these facts, we extract the $H-T$ diagram boundary $H_{c2}(T)$ from the $R(H,T)$ curves. Fig.~\ref{fig:figure_hc_tc} displays the results. Here, we have defined $H_{c2}$ as the value that gives 95\% of each case's normalizing resistance. This criterion ensures that the estimate is coherent with the resolution imposed by the noise level of our measurements. Meaningfulness relies on the fact that our measurements correspond to fields perpendicular to the surface of the film, and the possibility of surface superconductivity, i.e., the appearance of $H_{c3}$ may be ruled out. In fact, although the presence of such phenomena has been reported even for this configuration in Nb films \cite{kozhevnikov_17}, there is no indication of its prevalence in our data, i.e., $R$ presents a single sharp drop instead of a double-step structure. 

Finally, based on Fig.~\ref{fig:figure_hc_tc}, and given the celebrated Ginzburg-Landau relation
\begin{equation}
\label{eq:Hc2}
H_{\rm c2}=\frac{\Phi_{_0}}{2\pi\mu_{_0}\xi^2} \, ,
\end{equation}
we can conclude that the laser irradiation process gives way to (slight) changes in the fundamental properties of the superconductor. At a given temperature, the material's coherence length $\xi$ increases ($H_{\rm c2}$ decreases), and this takes place together with the reduction of the lattice parameter (Fig.~\ref{fig:figure_xrd_tc_gradual}).

\subsection{\label{sec:fluxpinning}Flux pinning landscape}

In the previous sections, we have dealt with the laser-induced variations of the intrinsic superconducting properties of Nb ($T_{\rm c}, H_{\rm c2}$). According to our observations, changes are systematic in terms of the laser pulse energy, and faintly dependent on the measurement configuration. In particular, differences related to the circulation of the transport current either parallel or perpendicular to the LIPSS (anisotropy) are hardly visible. Below, we will concentrate on the influence of laser irradiation in the (extrinsic) flux-pinning landscape. In particular, we will show that the behavior of the samples becomes anisotropic and also much sensitive to the experimental conditions (applied field and temperature). This study was performed at temperatures ${T\ge 6\,{\rm K}}$, in order to minimize the occurrence of magnetic flux avalanches, a detrimental phenomenon ubiquitous in films at low temperatures~\cite{colauto_21}, already investigated in Nb samples in our previous work~\cite{badia_24}.

\subsubsection{\label{sec:DM}Irreversible magnetization. Inductive $J_{\rm c}$}

As it is customary for type-II superconductors, we have analyzed the hysteretic magnetic response of the samples based on the isothermal $M(H)$ loops. {Considering that the thicknesses of the samples $d$ are affected by uncertainty (Fig.~\ref{fig:figure_sem_gradual}), we have preferred to evaluate the sheet current for each case, i.e., {$K_{\rm c}\equiv J_{\rm c}d$}. Note that the corresponding (approximate) $J_{\rm c}$ values are also displayed to ease comparison with literature. Recall that, according to critical state model (CSM) predictions, the induced critical current density is proportional to the width of hysteresis loops, $\Delta m$, which is obtained at each value of the applied field as the difference between the measured magnetic moments of descending ($m_\downarrow$) and ascending ($m_\uparrow$) field branches ($\Delta m = m_\downarrow - m_\uparrow$)}. For isotropic hard superconductors of square cross-section $2b\times 2b$ in longitudinal field, this is given by \cite{chen_90,brandt_95}:

\begin{equation}
\label{eq:CSMjc}
K_{\rm c} = \frac{3}{8}\frac{\Delta m}{b^3}  \; ; \qquad
J_{\rm c} = \frac{3}{8}\frac{\Delta m}{b^3d} = \frac{3}{2}\frac{\Delta M}{b}  \, .
\end{equation}
Here, we have introduced the sample's magnetization $\Delta M= \Delta m /V$,  $V$ the volume.
We note that the above analytical expression was obtained for superconductors with infinitely long geometry ($d\gg b$) and field applied parallel to the axis. Still, it is also generally valid for finite length when both branches are in fully penetrated states. In fact, it is important to recall that due to shape effects, the intrinsic $J_{c}(B)$ can be obtained from the width of the magnetization loop with the greatest accuracy in the case of thin films ($d\ll b$) with the field applied parallel to the shortest dimension \cite{sanchez_01}, as in this study (square thin films lying in the $xy-$plane with field applied along $z-$axis). Essentially, one may safely approximate $B\approx \mu_{_0}H$, i.e., self-field may be neglected. 

\begin{figure}[!]
\includegraphics[width=.45\textwidth]{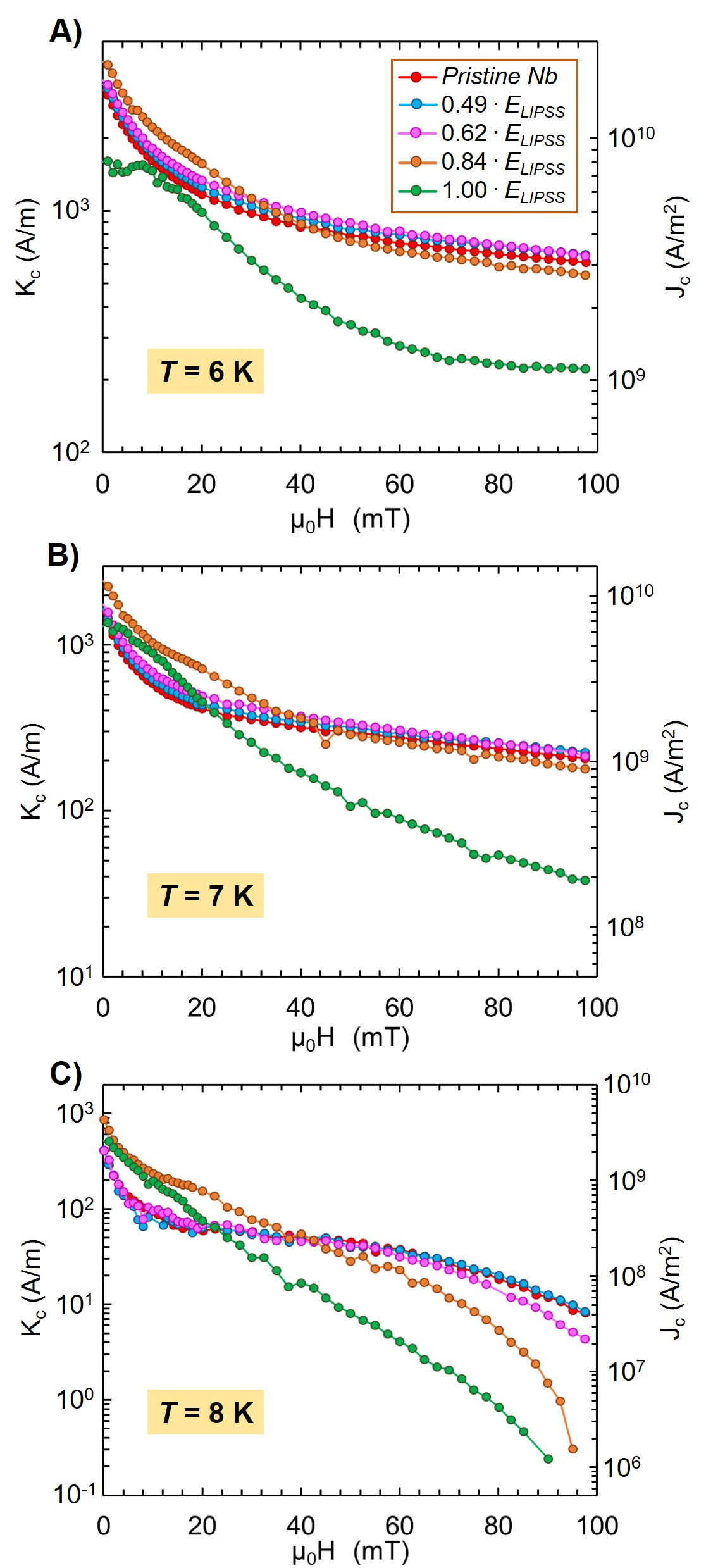}
\caption{\label{fig:figure_Jc_inductive} Field dependence of the inductive critical currents (Eq.~(\ref{eq:CSMjc})) at different temperatures: A) $T = 6K$, B) $T = 7K$ and C) $T = 8K$, for the square films under different irradiation conditions. The same samples (and color code) as in Fig.~\ref{fig:figure_xrd_tc_gradual}C are used here. }
\end{figure}

{Fig.~\ref{fig:figure_Jc_inductive} displays the induced $K_{c}$ and $J_{c}$ as a function of $\mu_{_0}H$ at 6, 7 and 8 K for the same square-shape films of Fig.~\ref{fig:figure_xrd_tc_gradual}C, as obtained by application of Eq.~(\ref{eq:CSMjc})}. These values were estimated from the third and fourth branches of the magnetization loops of the different samples, instead of using the first (initial) and second ones, so that the applied field fully penetrates the sample. In general, $J_{c}$ exhibits a temperature-dependent initial strong decay in the range $\mu_{_0}H< 10-30$~\unit{\milli\tesla}, followed by a smoother decay or ``plateau" and eventually a larger decay rate at higher fields. It must be remarked that the effect of laser irradiation with conditions below the threshold for the formation of LIPSS (${E_{\rm p}/E_{\rm LIPSS } < 0.7}$) is just a small but progressive increase of $J_{c}$ upon increasing laser pulse energy, being all these $J_{c}(\mu_{_0}H)$ curves very similar. On the contrary, once LIPSS start nucleating, strong variations on $J_{c}(\mu_{_0}H)$ are observed. Thus, compared to FS0, these samples show substantial improvements of $J_{c} $ at low fields ($\mu_{_0}H < 20-40 $ mT), mainly at 6~\unit{\kelvin} and 7~\unit{\kelvin}, whereas the opposite occurs at the highest fields.

As concerns the origin of the increased $J_{\rm c}$, a simplified interpretation was done in previous work~\cite{badia_24} by assuming a laser-induced anisotropy. Thus, according to the expression of the modified magnetization \cite{badia_24,gyorgy_89}:
\begin{equation}
M_{\rm aniso} =M_{\rm iso}\frac{3-1/\Gamma}{2}\, ,
\end{equation}
one may expect a growth of the saturation magnetic moment for the anisotropic (field-independent) case when the critical current density increases along the $x-$axis, while remaining as in the isotropic case in the $y-$axis (i.e. $J_{{\rm c}x}=\Gamma J_{{\rm c}y}$, with $\Gamma > 1 $ for the sample lying on the $xy-$plane and magnetic field applied along $z-$axis). 

{In passing, we note that a further argument aligned with the interpretation of our results in terms of anisotropy is the observation of bumps in the field dependence of the critical current density, which may be attributed to the appearance of flux-matching effects in the presence of a periodic potential related to the undulated topography of the sample \cite{martinoli_75}.} {In fact, using the expression $H_{1}\approx(\sqrt{3}/2)\Phi_{_0}\Lambda^2$ with $\Lambda$ the periodicity of the LIPSS structure mentioned before, one gets $H_1\approx 26.52$~\unit{\milli\tesla} in good agreement with the observed bumps in Fig.~\ref{fig:figure_Jc_inductive}.}

It must be stressed that the above considerations are just a {\itshape first approach} to the full problem. In particular, the relevance of the field dependence of $J_{\rm c}$ is masked by and mixed with the coexistence of current densities along the two anisotropy directions in the experiment. To remove ambiguity, we also performed transport measurements (Sec.~\ref{sec:transportJc}) in the bridge geometry so as to separate both effects.

Notwithstanding the above, the information provided by magnetization loops is very useful as it allows for analysis of the effect on the hysteretic behavior of gradually increasing the laser irradiation energy. Moreover, it gives access to the range of high current densities (${\sim 5\cdot10^8-1\cdot10^9~\unit{\ampere/\meter^2}}$), above the experimental scope in our transport measurements.

\subsubsection{\label{sec:transportJc}Anisotropic transport critical current density.}

To better establish the correlation between the physical properties with the circulation of currents along or transverse to the LIPSS, we measured the electrical resistivity of the three bridge samples (Fig.~\ref{fig:figure_sem_bridges}) for different applied transport currents. The use of high enough
current densities gives insight into the flux depinning dynamics, and thus, on the underlying pinning structures (recall the expression ${\bf F} = {\bf J}\times{\bf B}$ for the unit-volume depinning force). Primary experimental data, i.e., resistance curves for different values of the applied current, magnetic field, and temperature, additional explanations related to the aftermath of anisotropy in the analysis of transport data, as well as the influence of the threshold used to determine the critical current density are provided in Appendixes~\ref{app:A2},~\ref{app:A3}, and ~\ref{app:appB} respectively.

\begin{figure}[!]
\includegraphics[width=0.45\textwidth]{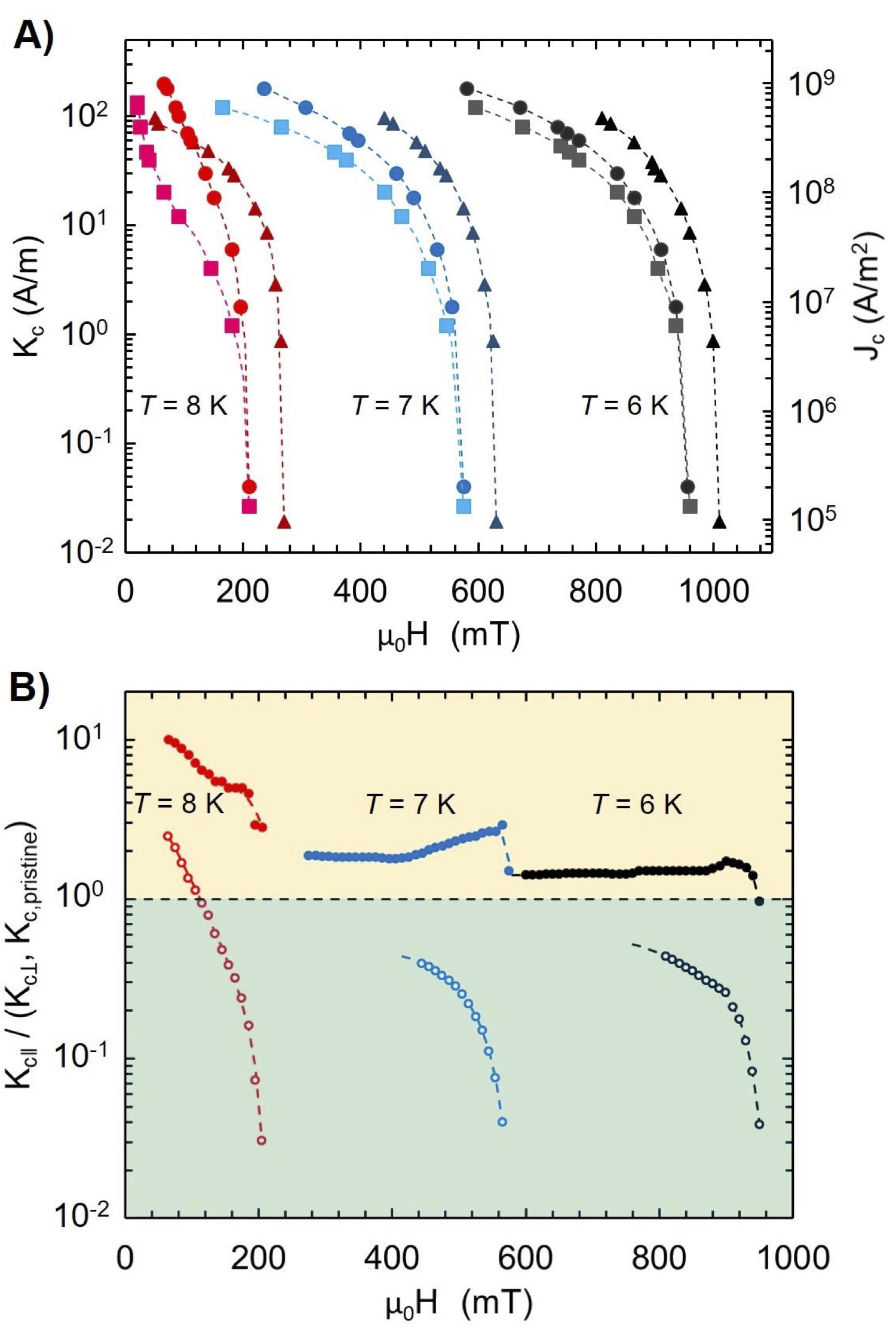}
\caption{\label{fig:figure_jch_bridges_aniso_factors} A) Critical sheet currents ($K_{\rm c}\equiv I_{\rm c}/W$) obtained from transport measurements for the pristine and laser-treated samples. Triangles correspond to the untreated Nb bridge ($K_{\rm c,pristine}$), circles are for the parallel configuration bridge ($K_{{\rm c}\parallel}$), and squares for the perpendicular configuration ($K_{{\rm c}\perp}$).
B) Field dependence of the anisotropy ratios $K_{{\rm c}\parallel}/K_{{\rm c}\perp}$ (full symbols) and $K_{{\rm c}\parallel}/K_{\rm c,pristine}$ (open symbols) of the transport critical sheet currents obtained for the "parallel", "perpendicular" and "pristine" bridge configurations at the temperatures of 6, 7 and 8~\unit{\kelvin}. The horizontal line (unit ratio) separates the regimes for which the laser treatment either improves or deteriorates the transport properties of the films, i.e., $K_{{\rm c}\parallel}\gtrless K_{\rm c,pristine}$.}
\end{figure}

In Fig.~\ref{fig:figure_jch_bridges_aniso_factors}A, the comparison of the critical current densities derived for the two irradiated bridge samples systematically shows higher/lower values of $J_{\rm c}$ for currents flowing along/across the LIPSS. Also, at each temperature, a tendency to the convergence of both curves for the higher values of the applied field is observed. We notice that this ``high-field and low-current limit'' of the $J_{\rm c}(H)$ curves may be identified with $H_{c2}$ for the sample at the given temperature. The convergence of the values for the parallel and perpendicular orientations agrees with the discussion in Sec.~\ref{sec:ht}, i.e., the orientation-independent modification of the superconducting reversible properties by laser-irradiation.

Thus, as $H$ approaches $H_{\rm c2}$ (which corresponds to the measurements for the lowest transport currents in Fig.~\ref{fig:figure_jch_bridges_aniso_factors}), one may note the downshift in the $H-T$ line in Fig.~\ref{fig:figure_hc_tc}. On the other hand, at 8~\unit{\kelvin}, a crossover between the curves corresponding to the samples BS0 and BS$_{\rm PAR}$ samples is observed. The irradiated sample BS$_{\rm PAR}$ shows higher performance. Thus, aligned with the observation for the inductive critical currents (Fig.~\ref{fig:figure_Jc_inductive}), the transport critical current density values of the laser-irradiated samples are higher for the higher-temperature \& lower-field regime. Strictly speaking, owing to our experimental limitations, the crossover is observed for the parallel orientation and guessed for the perpendicular case. It is remarkable that the crossover current density values at 8~\unit{\kelvin} derived from both techniques are basically coincident $K_{\rm c}\approx 60$~\unit{\ampere/\meter}. It is apparent, however, that the value of the applied magnetic field at which the crossover occurs in the two measurements differs noticeably. In our view, the reduction of this field for the inductive measurements may be related to the indivisible appearance of the components $K_{\rm c\parallel}$ and $K_{\rm c\perp}$ in such a case.
Finally, we note that the characteristic $J_{\rm c}(H)$ decay displayed by the inductive critical current densities, i.e., initial sharp drop followed by a smoother rate for higher fields, is also observed in the transport-derived critical current densities.

For further clarity concerning the anisotropic nature of the critical current densities,  Fig.~\ref{fig:figure_jch_bridges_aniso_factors}B displays the current density ratios as a function of the magnetic field at the different temperatures investigated. The plot splits into two regions to highlight the cases for which $K_{\rm c\parallel}$ surpasses, equals, or is lower than $K_{\rm c\perp}$ or the pristine $K_{\rm c}$ values. The following tendencies are observed: (i) $K_{\rm c\parallel}$ and $K_{\rm c\perp}$ display nearly field-independent ratio ($>1$) for some ranges of the magnetic field, and (ii) eventually collapse to equality for fields approaching $H_{c2}$. (iii) The critical current density perpendicular to the LIPSS, $K_{\rm c\perp}$, is generally the lowest, with a tendency (not fully resolved with our data) to exceed the pristine value for their lower field values at the temperature of 8~\unit{\kelvin}. Lastly (iv), we stress that, unless for the highest temperatures and lowest applied fields, the laser-irradiated samples display reduced flux pinning capabilities as indicated by the ratio $K_{\rm c\parallel}/K_{\rm c,pristine}$.

\subsubsection{\label{sec:moi}Anisotropic critical current density from MOI}

A dedicated characterization of selected square-shaped films irradiated with different pulse energies was carried out by MOI experiments. This technique allows direct visualization of the anisotropy of $J_{\rm c}$ in superconductors.

\begin{figure*}[t]
\includegraphics[width=.9\textwidth]{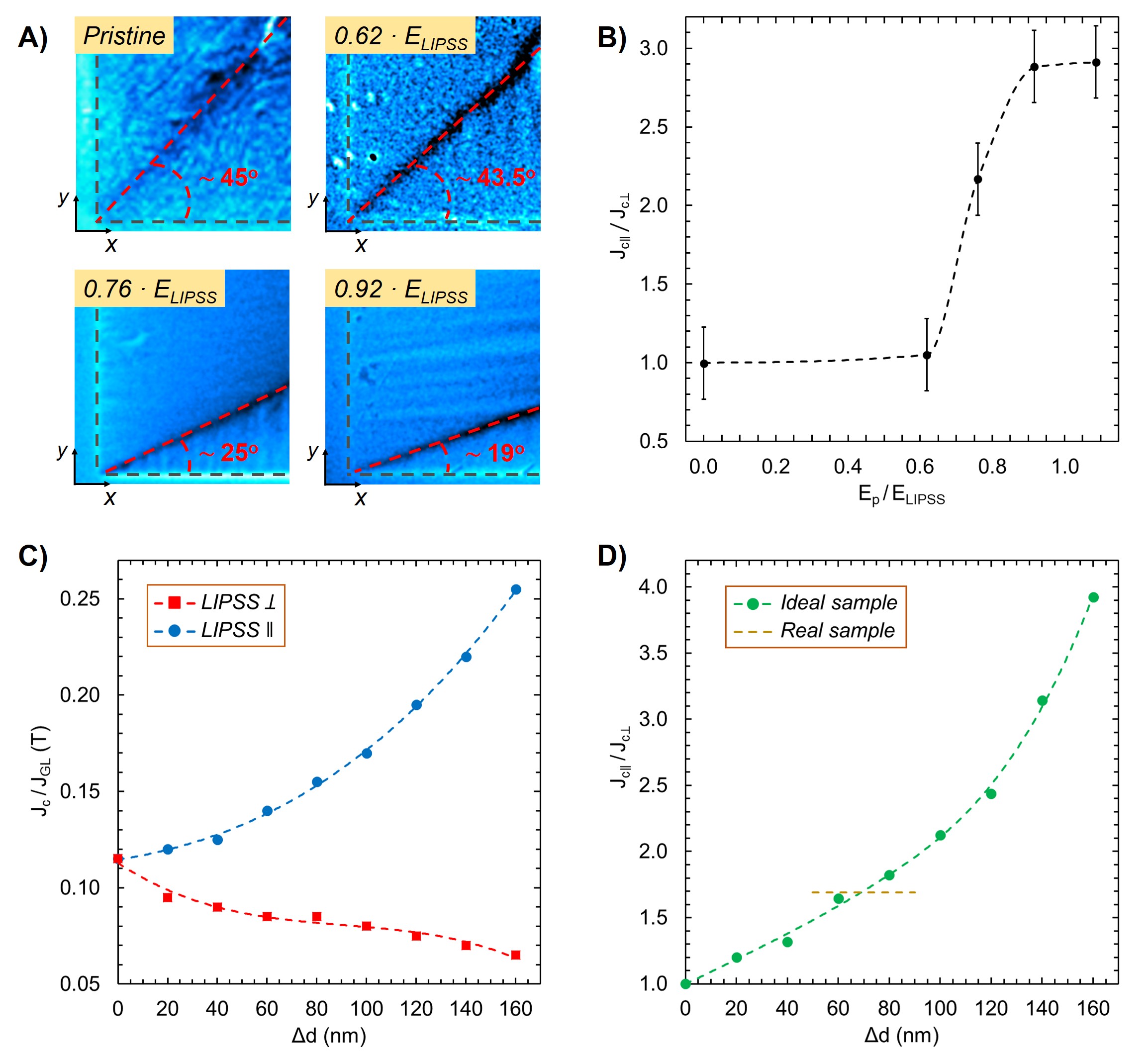}         
\caption{\label{fig:figure_MOI}{A) Magneto-optical images of the corner of the pristine and laser-treated samples with different pulse energies at $T/T_{\rm c} \approx 0.85$. Dashed grey and red lines highlight the sample's edges and the $d-lines$ (see text), respectively. The average $\alpha$ values, considering all four samples' $d-lines$, are given in each image. (B) $J_{{\rm c}\parallel}/J_{{\rm c}\perp}$ ratio obtained from MOI measurements employing Eq.~(\ref{eq:anisoMOI}), as a function of the relative laser pulse energy, $E_p/E_{LIPSS}$. The lower panels provide TDGL simulation results: (C) $J_{{\rm c}\parallel}$ and $J_{{\rm c}\perp}$ for an ideal undulated thin film of dimensions {$5~\mu{\rm m}\times\, 5~\mu{\rm m}$} (and thickness 200~nm), in applied magnetic field $\mu_0 H=4$~mT and $T/T_{\rm c} \approx 0.85$, in terms of the undulation's depth. (D) The anisotropy ratio predicted for the ideal sample, stemming from (C). Dashed lines connecting the symbols are guides for the eye. For comparison, the horizontal dashed line in (D) gives the ratio obtained for the profile of a real sample with $\Delta{\rm }d = 70\pm 20$~\unit{\nano\meter} (see text).}}
\end{figure*}

Figure \ref{fig:figure_MOI}A displays the corners of selected samples obtained by MOI after zero-field cooling the samples to a set temperature and gradually increasing the magnetic field. The temperature and field values in the figure were chosen so as to optimize the visualization and to highlight the differences among samples. For these pictures, the laser polarization in the irradiation process was parallel to the $y-$axis, so that LIPSS would eventually be generated parallel to the $x-$axis. Concerning the interpretation of Fig.~\ref{fig:figure_MOI}A, we recall that in flat samples subjected to applied magnetic fields in the $z-$direction, the current density streamlines induced in the $x-y$ plane bend sharply, defining the so-called $d-lines$ with a characteristic angle, $\alpha$, that depends on the anisotropy in the plane \cite{schuster_94,jooss_02} and is clearly visible in the MOI experiments as dark regions where the screening of the applied field is at its highest. In particular, for the images displayed in Fig.~\ref{fig:figure_MOI}A, 
\begin{equation}
\label{eq:anisoMOI}
{\rm tan}\,\alpha = J_{{\rm c}y}/J_{{\rm c}x}  = J_{{\rm c}\perp}/J_{{\rm c}\parallel}\equiv\frac{1}{\Gamma} \, .
\end{equation}

As expected, the magnetic flux penetrates isotropically for the pristine film, i.e. $\alpha  \approx $ 45$^{\circ}$. However, as the laser pulse energy increases above approximately ${0.60\cdot E_{LIPSS}}$, a sharp decrease in $\alpha$ is observed. Note that this energy corresponds to the onset of LIPSS formation (Figure \ref{fig:figure_sem_gradual}) and the onset of the lattice parameter reduction (Figure \ref{fig:figure_xrd_tc_gradual}B). More specifically, the angle reduces to about 25$^{\circ}$ in the sample irradiated with ${0.76\cdot E_{LIPSS}}$, and further to  19$^{\circ}$, for the sample irradiated at ${0.92\cdot E_{LIPSS}}$, with almost fully developed LIPSS. 
The variation of the anisotropy parameter $\Gamma$ obtained by application of Eq.~(\ref{eq:anisoMOI}) is shown in Fig.~\ref{fig:figure_MOI}B. As the corresponding MOI measurements were performed using a full magnetic cycle loop of ${\pm 4}$~\unit{\milli\tesla}, one may infer that the observed anisotropic effects emerge already at low fields.

\subsubsection{\label{sec:tdgl}Anisotropic critical current density from TDGL}

Here we put forward the results of our TDGL simulations, targeted to understand the role played by the surface topography on the anisotropy of the critical current density. As already noticed in early experimental work~\cite{martinoli_75,niessen_65, morrison_70,daldini_74}, based on the concept of {\itshape vortex line energy} (free energy per vortex unit length), one may expect that undulations serve to ``pin vortices'' as they would prefer to sit at the valleys and require energy for hopping across the undulations (which corresponds to the {\itshape force} component given by $\bm{J}_\parallel\times \bm{\Phi}_{_0}$). Also, the deeper the structure, the higher the $J_{\rm c\parallel}$ value as compared to the current necessary to make vortices slide along the valleys ($J_{\rm c\perp}$), going to zero in an ideal crystal. The above oversimplified picture neglects realistic effects, such as background pinning, edge barriers, vortex-vortex interactions, and imperfections in the undulating structure. Nevertheless, as shown below, the monotonic increase of anisotropy with the depth of the grooves is not only an experimental observation, but also a prediction of the TDGL simulations under logical assumptions, and with reasonable quantitative comparison to experiments.

Fig.~\ref{fig:figure_MOI} (panels C, D) shows the results obtained for an idealized variable thickness profile given by ${d(x,y) = d_0-(\Delta d/2)\left[1+\cos(2\pi x/L)\right]}$, with ${d_0 = 200~{\rm nm}}$, $L = 250$~nm, and $\Delta d$ as a free parameter that determines the depth of the undulation. The values of these parameters mimic the experimental situation in our series of laser-irradiated samples. Also, we show the current density anisotropy obtained for a real thickness profile, as derived from AFM measurements (see Ref.~\cite{martinez_25}), used as exact $d(x,y)$ input in the simulations. In all simulations, the coherence length is $\xi(T) = 25$~nm, corresponding to temperature $T \approx 7.35$~K and critical temperature $T_{\rm c} = 8.65$~K.
For these instances, we take a square superconductor of finite dimensions ${5~\mu{\rm m} \times\, 5~\mu{\rm m}}$, in applied magnetic field $\mu_0 H = 4$~mT. The critical current densities were defined by the onset of net vortex motion, in this case, also overcoming the barrier at the sample edges. We note that the anisotropy ratio $J_{c\parallel}/J_{c\perp}$ grows as the undulations are made deeper (i.e. with $\Delta d$ increasing), in reasonable agreement with experimental data though the latter show the tendency of saturation. In fact, one should take into account that the plateaus in Fig.~\ref{fig:figure_MOI}B are nothing but the thresholds for the nucleation of LIPSS and the eventual perforation of the film, i.e.: within the experimental range of LIPSS formation one has indeed a monotonic increase of anisotropy.

According to the data in \cite{martinez_25}, the depth of undulations for the optimized sample is $\lesssim 80$~\unit{\nano\meter}, which means that the experimental anisotropy values are a bit over simulations. However, considering the assumptions of the model, the quantitative agreement is noticeable. Related to this, the ideal periodicity ansatz may be considered reliable. In fact, as shown in Fig.~\ref{fig:figure_MOI}D, by using the actual AFM scan profile of the reference sample, where the undulations are not ideally periodic, we confirm that the anisotropy ratio is validly determined by the dominating undulation depth ($\approx 70\pm 20$~\unit{\nano\meter} in the selected region). 

\begin{figure}[!]
\includegraphics[width=0.45\textwidth]{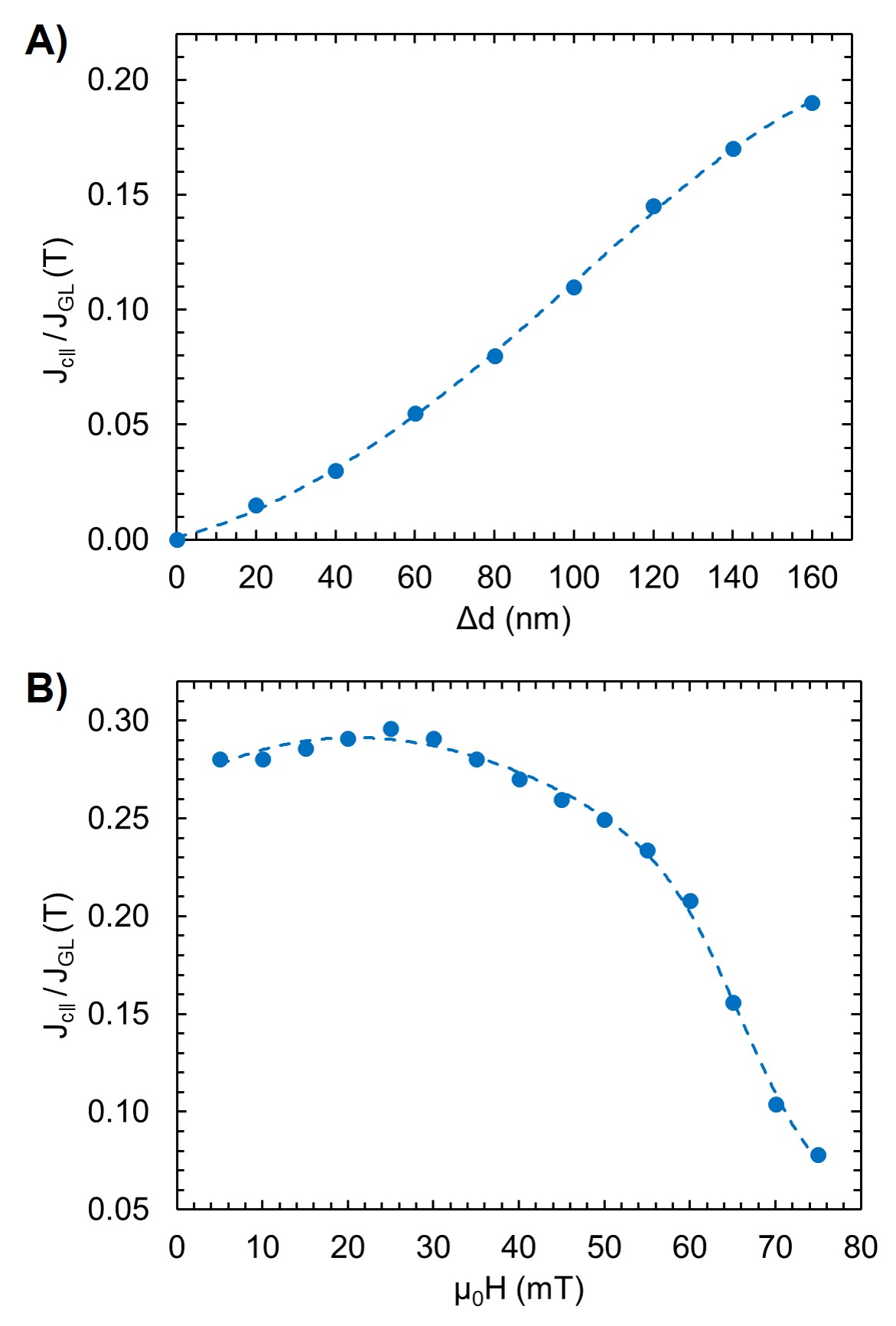}
\caption{\label{fig:simuJcH} TDGL simulations for an infinite (periodic) ideally undulated sample. (A) $J_{c\parallel}$ obtained from simulations in applied magnetic field ${\mu_0 H = 4~{\rm mT}}$, as a function of the depth of thickness undulations, $\Delta d$. (B) $J_{c\parallel}$ obtained from TDGL simulations on an infinite (periodic) sample, with thickness undulated by $\Delta d=100$~nm, as a function of applied magnetic field.
Dimensionless units are used for $J_{\rm c}$  in terms of the temperature-dependent Ginzburg-Landau critical current density.}
\end{figure}

In order to clearly detach the pure contribution of the undulations to the critical current and eliminate the effect of the edges, we next calculated the vortex dynamics in a periodic (infinite) sample, at the same magnetic field and temperature (Fig.~\ref{fig:simuJcH}). In such a case, $J_{c\perp}=0$, since there is no pinning in the idealized theoretical sample to prevent the vortex motion along the thickness grooves. On the other hand, $J_{c\parallel}$ grows with $\Delta d$, as growing undulations increase the effective pinning potential that hampers vortex motion. We note that the $J_{c\parallel}$ obtained in this case exactly corresponds to the difference $J_{c\parallel}-J_{c\perp}$ seen in the case of a finite sample (Fig.~\ref{fig:figure_MOI}C), and thus indeed represents the contribution of the undulations alone.

Finally, to further understand the vortex behavior in the periodically undulated thin film, we fixed the undulations at $\Delta d=100$~nm, and varied the applied magnetic field (i.e., the vortex density). This allows one to comprehend the experimental observation of a non-monotonic dependence of the undulation-related contribution to $J_{c\parallel}$ on magnetic field, increasing at low fields and decreasing at higher fields (see Fig.~\ref{fig:simuJcH}B).
Such a behavior stems from the fact that, initially (and mediated by vortex-vortex interactions), the increasing vortex density stabilizes the lattice with vortices occupying the floor of the undulations. This works against their mobility across the grooves until eventual hopping across the grooves occurs in the form of vortex bundles. Once vortices sufficiently populate the undulations, the vortex lattice no longer drifts collectively, and the vortex dynamics primarily consists of individual vortex hoppings between the thickness grooves. Those hoppings are easier to stimulate by applied current as the vortex density is increased, and then $J_{\rm c}$ decreases.
The transition from a {\itshape collective} to {\itshape individual} vortex drift regime may be clearly discerned through the simulated vortex dynamics for different values of the applied magnetic field, as shown in the videos provided in the supplementary material (see also Appendix~\ref{app:appC} for explanation). There, we illustrate the flux flow regime for the magnetic flux densities $\mu_{_0}H=25$~\unit{\milli\tesla} and $\mu_{_0}H=65$~\unit{\milli\tesla}, corresponding respectively to the maximum of $J_{\rm c\parallel}(H)$ in Fig.~\ref{fig:simuJcH}B and to a field well above such value.
 
\section{\label{sec:conclusions}Conclusions}

Ultraviolet femtosecond pulsed-laser-irradiation (${\lambda =343\, {\rm nm},\tau_p={238}\, {\rm fs}}$) has enabled the formation of surface nanostructures on Nb thin films, which gradually evolve to quasi-1D periodic corrugations (LIPSS) of lateral periodicity $\Lambda\approx 260$~\unit{\nano\meter}. Our study includes a systematic characterization of a series of polycrystalline Nb thin films in successive stages of the LIPSS formation process, starting from the pristine sample and covering the full range up to the perforation of the film. In practice, a series of 12 formation steps was generated by tuning up the laser pulse energy while keeping other inputs constant. Square films as well as micro-sized bridges with LIPSS either parallel or perpendicular to the path (and also without LIPSS) were fabricated.   

Some fundamental questions have been studied: (i) the influence of the laser irradiation conditions on the intrinsic superconducting properties, i.e.: critical temperature and upper critical field; (ii) the evolution from isotropic to anisotropic behavior of the critical current and its dependence on the applied magnetic field and temperature; (iii) the correlation between the laser-processing parameters and the transformations of microstructural and physical properties.
A direct correlation has been established between the progressive formation of LIPSS and the underlying microstructural changes. The onset of LIPSS formation also triggers: (i) a reduction of the film thickness, pointing towards a certain level of ablation, in agreement with recent work in stainless steel~\cite{Wonneberger_25}; (ii) a systematic reduction of the lattice parameter $a$ of Nb; (iii) a small increase of the grain size mainly at the LIPSS' peaks. A moderate variation of the intrinsic superconducting properties accompanies these features. In fact, a well-defined downshift of the $H_{\rm c2}(T)$ line occurs. Specifically, this reduction is characterized by the values $\Delta T_{\rm c}\approx 0.3$~\unit{\kelvin} and $\Delta H_{\rm c2}\approx 50$~\unit{\milli\tesla} for films with fully formed LIPSS, and occurs isotropically. The similarities between $T_{\rm c} (E_{\rm p})$ and $a(E_{\rm p})$ dependencies suggest a correlation between both parameters. {Our work complements previous studies~\cite{bose_05,liu_09}. There, the experimental dependence of $T_{\rm c}$ on the value of the lattice parameter was reported, showing that the expansion of the lattice parameter beyond the reference value $a=3.31~\textup{\AA}$~\cite{XRD} is also accompanied by a reduction of $T_{\rm c}$. On the other hand, in recent work~\cite{choi_20}, the authors discuss that, in fact, $T_{\rm c}$ is reduced both by compressive and tensile stresses, which corroborates the full picture.
Furthermore, the observable lattice compression found for samples fully covered by LIPSS (${\approx 1.05\%}$, as compared to the pristine reference) is likely to produce stress in the surrounding areas when the sample is irradiated in localized zones.
Based on the results obtained in this work, we conclude that this may be the reason for the important decrease of $J_{\rm c}$ at the boundaries between LIPSS and pristine neighboring areas observed in previous studies~\cite{martinez_25}.}

The flux pinning-related properties, specifically the critical current density, are noticeably affected by the laser irradiation process. Changes occur gradually as LIPSS are being formed, anisotropically as concerns the relative orientation of the applied current in experiments, and differently for different operation conditions (applied field and temperature). The analysis of magnetic hysteresis and transport measurements reveals that optimally treated samples possess higher critical current densities in a specific range of field and temperature ($H\lesssim 100$~\unit{\milli\tesla}, $6$~\unit{\kelvin}$\,\lesssim T\lesssim$\,8\unit{\kelvin}) as compared to the pristine film. The increase in critical current density is related to the circulation of currents along the direction of the LIPSS and occurs progressively in parallel to their formation process, as deduced from transport experiments and magneto-optical imaging. A complex dependence of the anisotropy factor $J_{\rm c\parallel}/J_{\rm c\perp}\equiv \Gamma (H,T)$ is observed, with general trends indicating an increase with temperature, a saturation for the lower applied magnetic fields, and a collapse to unity (isotropy) when the sample approaches the reversible regime, i.e.: $H\to H_{\rm c2}$.

Our findings concerning the induced anisotropic behavior of the laser-irradiated samples have been interpreted in terms of the phenomenological critical state theory, whose central concept is the macroscopic critical current density $J_{\rm c}$ (sub-mm scale) and further supported by mesoscopic scale calculations (sub-$\mu$m scale) of the vortex dynamics in undulated samples through the time-dependent Ginzburg-Landau theory. According to both the experimental observations and the simulations, the presence of thickness variations of depth $\lesssim 100$~\unit{\nano\metre} in our films with original thickness of $200$~\unit{\nano\metre} produces an effective flux pinning effect that multiplies the critical current density along the channels by a factor of $\approx 2$, related to the preference of vortices to settle in the lower thickness region. Some features of the complex $J_{\rm c}(H,T)$ behavior have been explained within the TDGL framework. In particular, the non-monotonous dependence of the undulation-related $J_{\rm c\parallel}(H)$ aligns with the theoretical observation of the stabilization of the vortex lattice with increasing densities in the vortex chains within the channels defined by the LIPSS.
%

In summary, this work delves into the nature of the laser-induced modification of the physical properties of superconducting Nb films. \abm{We demonstrate control over the radiation parameters that allow the modification of the physical properties {\itshape {\`a} la carte}.} In fact, we find a correlation between the variation of the fundamental physical parameters (critical temperature, upper critical field, critical current density, and the appearance of anisotropy) with the actual material processing conditions. \abm{These results qualify laser-processing as an enabling technology in the fabrication of superconducting circuit elements. Investigation is underway to achieve further control of the surface profiles, including the fabrication of inclined ripple nanostructures, as required in superconducting filters, diodes, and field-resilient ratchets. Although this work focuses on Nb films, the technology may be transferred to other materials, and work is in progress for type-II alloys and high-T$_{\rm c}$ compounds.}

\section*{Acknowledgments}
This publication is part of the projects PID2020-113034RB-I00 (funded by MCIN/AEI/ 10.13039/501100011033), PID2023-146041OB-C21 (funded by {MICIU/AEI/10.13039/501100011033} and ERDF/EU) and T54-23R (funded by Gobierno de Aragón).  J.F. acknowledges support from Gobierno de Aragón through predoctoral contracts. N.L. acknowledges the support from FRS-FNRS (Research Fellowships
FRIA). The work of E.F. has been financially supported by
the FWO and F.R.S.-FNRS under the Excellence of Science (EOS)
Project No. O.0028.22. The authors would like to acknowledge the use of Servicio General de Apoyo a la Investigación-SAI (Universidad de Zaragoza) and the Spanish National Facility ELECMI ICTS, node ``Laboratorio de Microscopías Avanzadas (LMA)'' at ``Universidad de Zaragoza''. The work of L.R.C. and M.V.M. was sponsored by the Army Research Office and was accomplished under Grant Number W911NF-24-1-0145. The views and conclusions contained in this document are those of the authors and should not be interpreted as representing the official policies, either expressed or implied, of the Army Research Office or the U.S. Government. The U.S. Government is authorized to reproduce and distribute reprints for Government purposes notwithstanding any copyright notation herein. The collaboration in this work has been fostered by the EU-COST Action CA21144 SUPERQUMAP.

\section*{Data availability}
The data that support the findings of this article are openly available at 
\cite{zenodo_frechilla_25}.

\appendix
\renewcommand\thefigure{\thesection.\arabic{figure}}    

\appsection{\label{app:appA} Resistive behavior: experimental features}
\setcounter{figure}{0} 

Below, we provide additional information on the measured $\rho (H,T)$ curves that  were used to obtain the superconducting parameters of the films presented in the main text. This includes measurements for low transport current densities, that reveal the intrinsic parameters $T_{\rm c}$ and $H_{\rm c_2}$, as well as higher transport currents, probing the flux pinning regime and providing the critical current density $J_{\rm c}(H, T)$. Also, we will provide the analysis of an extrinsic experimental feature: the appearance of a characteristic peak effect in some standard experimental configurations. The methodology for the deconvolution of the data in such cases is presented.

\begin{figure*}[!]
\includegraphics[width=.85\textwidth]{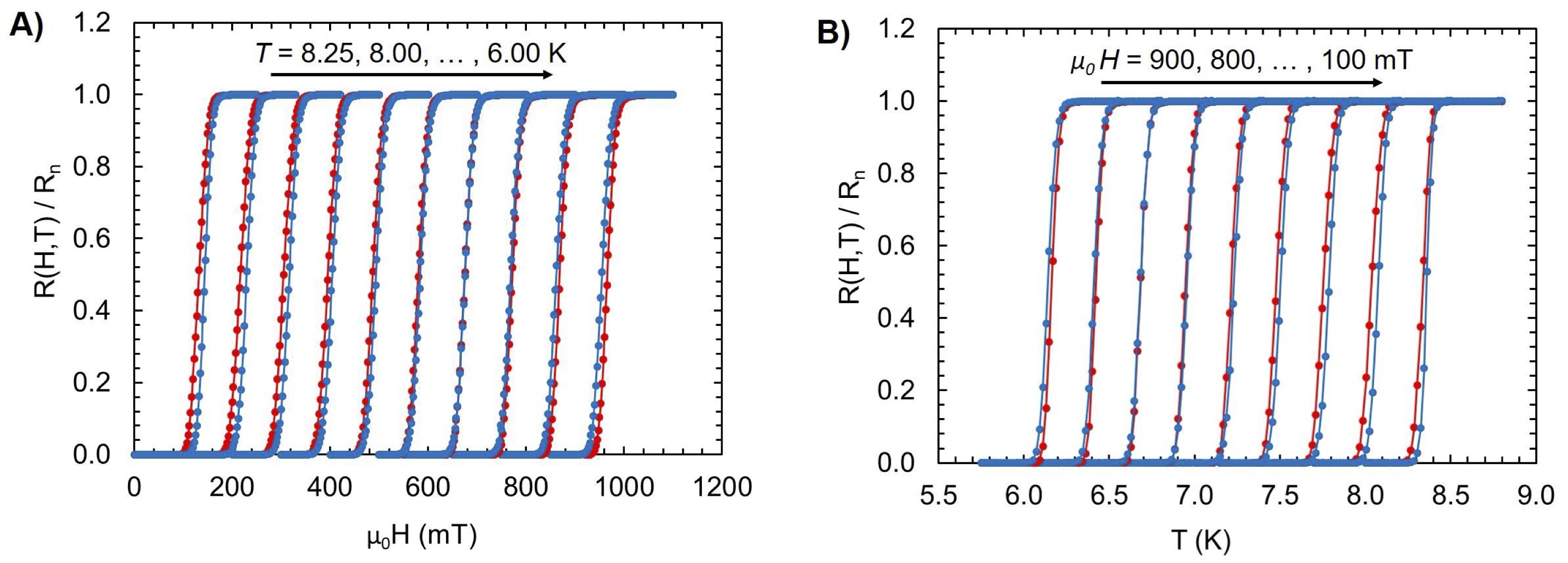}
\caption{\label{fig:figure_hc_tc_low} Normalized resistance of the Nb bridges BS$_{\rm PAR}$ and BS$_{\rm PERP}$ for applied current {$I=45\,$\textmu{A}}. For each case, $R_{\rm n}$ stands for the value $R(T=10$~\unit{\kelvin}). Panel A) shows isothermal magnetoresistance curves (from 8.25 to 6 K in steps of 0.25 K), B) displays the resistance vs. temperature for a collection of values of the applied magnetic field (in steps of $100$~\unit{\milli\tesla}).
} 
\end{figure*}

\begin{figure*}[!]
\centering
\includegraphics[width=.95\textwidth]{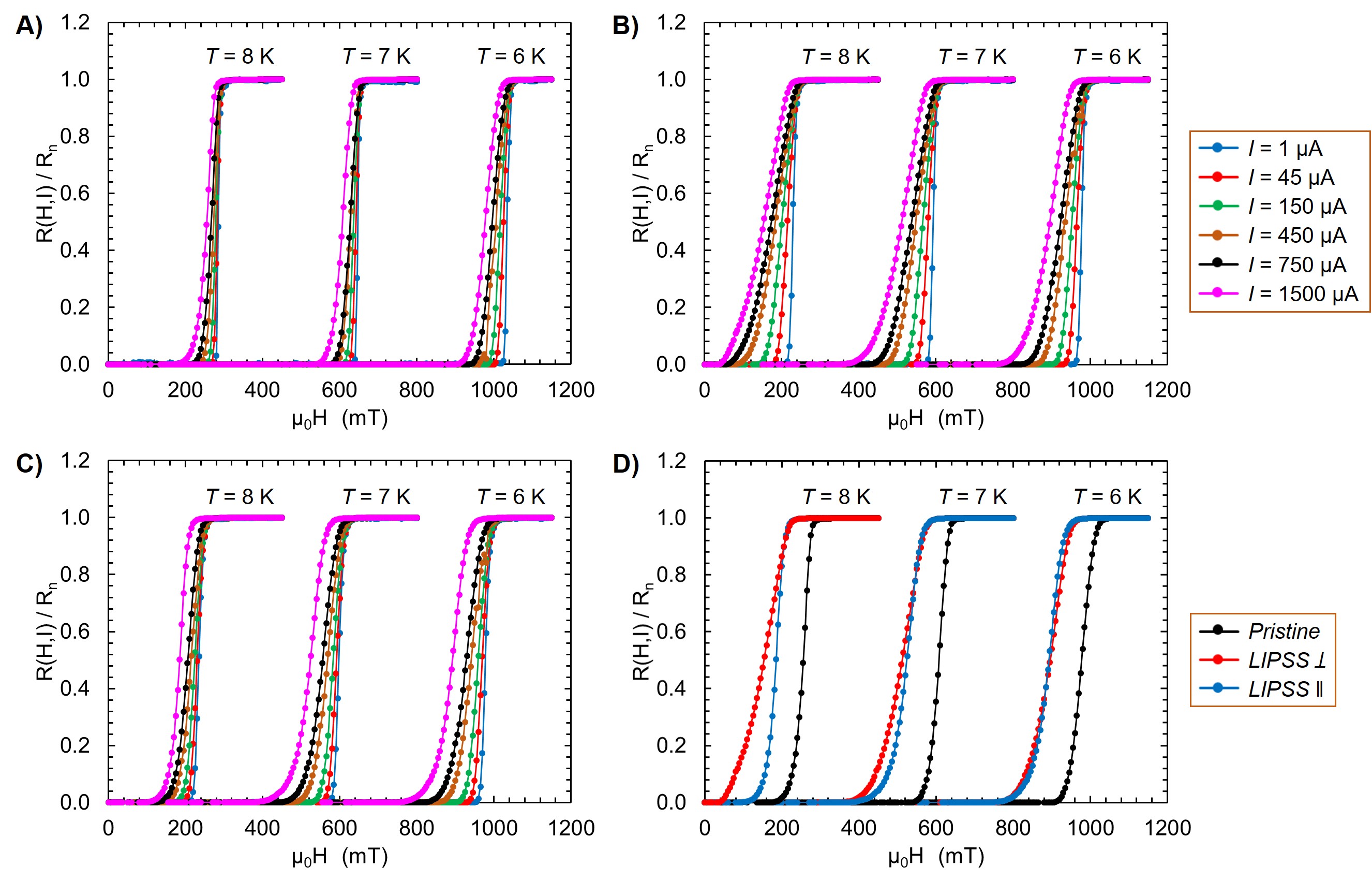}
\caption{\label{fig:figure_rht_bridges} Normalized resistance of bridge samples as a function of field at different temperatures and transport currents: A) pristine (BS0), B) LIPSS $\perp$ to transport current (BS$_{\rm PERP}$) and C) LIPSS $\parallel$ to transport current (BS$_{\rm PAR}$). Panel D) displays the transition at $I$ = 1500 $\mu$A for the three samples to ease comparison. The upper label-box corresponds to panels A)-C) and the lower one to panel D).}
\end{figure*}

\subsection{\label{sec:lowJ} Low current densities: reversible regime}

Taking advantage of the well-defined plateau beyond the rather steep transition for small transport current values ($I=45$~\unit{\micro\ampere} in this case), we consider the normalized resistance relative to the value at the plateau $R_n$. 
 As it is clear in  Fig.~\ref{fig:figure_hc_tc_low}A and \ref{fig:figure_hc_tc_low}B, the resistance curves do not show evidence of relevant broadening in this range. In fact, the transitions at changing fields and temperatures are nearly a translation one from the other, just with a subtle crossing effect between the curves for the two orientations. Note, for instance, that the $R(H,T_{_0}={\rm constant})$ curves for the perpendicular orientation (as compared to the parallel case) shift to lower fields at the lower temperatures, crossing to higher fields as $T$ increases. In other words, at lower temperatures $H_{\rm c2}$ is slightly higher for the parallel orientation and vice versa. We note in passing that these subtle differences in the resistivities either along or across the LIPSS may also be detected in the less straightforward van der Pauw configuration for square films. In fact, as shown in Sec.~\ref{app:A3}, the 2D anisotropic configuration is marked by a characteristic peak, whose width parameterizes the differences between the two orientations.

On the other hand, related to the fact that hysteresis effects were not detected when comparing forward and backward measurements, we can consider that the curves in Fig.~\ref{fig:figure_hc_tc_low} are representative of the equilibrium phase diagram of the samples. {This will be further supported by the quantitative analysis performed in Sec.~\ref{app:appB}. In brief, for the level of current used in these experiments, the obtained {\itshape critical field} is basically independent of the actual values of the current and the criterion used to define the transition. This excludes the impact of extrinsic effects such as those coming from the flux pinning landscape.}

\subsection{\label{app:A2} High current densities: irreversible regime}

Results are displayed in Fig.~\ref{fig:figure_rht_bridges} for the temperatures of 6, 7, and 8~K under magnetic fields perpendicular to the film. In this figure, we observe a noticeable broadening of the transition upon increasing the applied current, contrary to the case of Fig.~\ref{fig:figure_hc_tc_low} (changing $H_{_0}, T$ in the low current limit). In particular, this effect is more pronounced
for the bridge BSper. Notice that, again, taking advantage of the well-defined plateau above the transition, the curves are normalized to the value $R_n$ for each sample.

The experimental data in Fig.~\ref{fig:figure_rht_bridges} were used to extract the critical current densities as a function of the applied magnetic field. Some comments are due. Contrary to the common practice of deriving the critical current from current-voltage curves, here we identify the given applied current values as the critical currents for a certain applied field and temperature, i.e.: $I_{\rm c} (H,T)$  through the condition of reaching a threshold resistance value. In particular, Fig.7A in the main text, is {obtained by using $R_{\rm thres}=0.01 R_{\rm n}$ for each sample} and $(H,T)$ pair.
Other criteria, such as the use of a constant electric field (for instance, the customary $E_{\rm c} = 1$~\unit{\micro\volt/\centi\meter}) have been explored in order to assess that our conclusions are independent of the criterion. This technical issue is discussed in Appendix \ref{app:appB}. 

\begin{figure}[!]
\includegraphics[width=.45\textwidth]{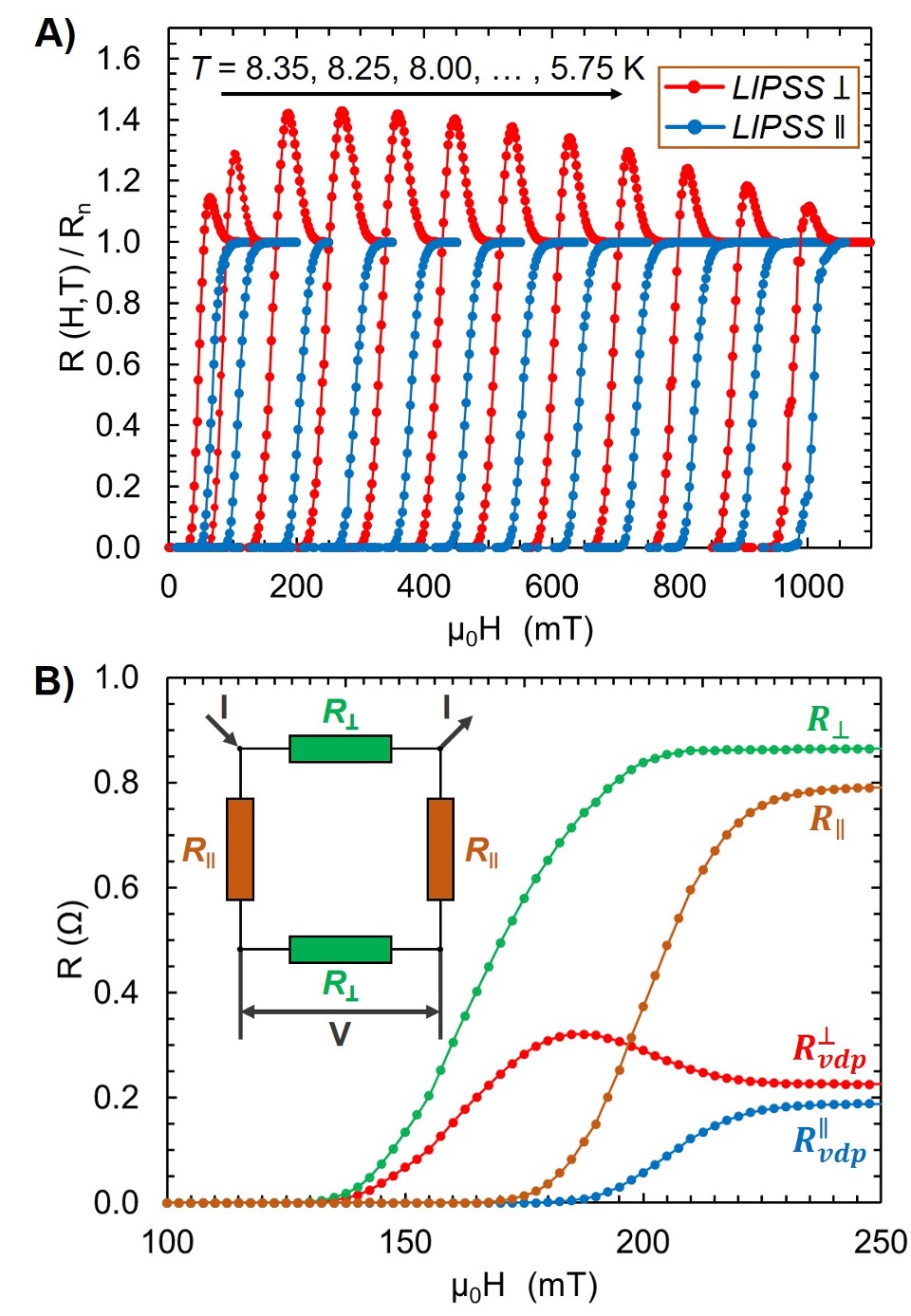}     
\caption{\label{fig:figure_A1}{A: Experimental normalized resistive transitions obtained for the square thin film FSL ($I_{\rm dc} =$ 4.5 mA}) in the van der Pauw configuration, with current injected either parallel or perpendicular to the LIPSS direction. B: Application of the lumped model (inset) to invert the resistive data of the anisotropic thin film. Experimental data correspond to the curves for $T=8$~ K in panel A.}
\end{figure}

\subsection{\label{app:A3} Anisotropy: the peak effect property}

Transport experiments in superconducting samples may display an ``extrinsic'' peak issue related to the interplay between anisotropy and geometric finite-size effects concomitant with the measurement method~\cite{Buzea_2000, Buzea_2001}.
Specifically, in our case, we performed transport measurements in square thin films, using the van der Pauw configuration as sketched in Fig.~\ref{fig:figure_A1}, obtaining results such as those shown in the left panel. In this figure, we notice the appearance of resistance peaks noticeably exceeding the normal-state plateau value $R_n$, when measurements are performed by injecting the transport current through the pads perpendicular to the LIPSS (inset in Fig.~\ref{fig:figure_A1}B). In contrast, the peaks disappear if measurements are performed when the current is injected through the pads in the parallel direction. Furthermore, they are never observed for the case of thin bridges with LIPSS aligned either along the channel or transverse to it.
As already discussed in previous studies,~\cite{Buzea_2000, Buzea_2001}, this effect may be attributed to anisotropy in the critical superconducting parameters, i.e., \(T_{\rm c}\) or \(J_{\rm c}\) together with the configuration of the transport measurements.

The central concept behind the above arguments may be illustrated in terms of the toy model in Fig.~\ref{fig:figure_A1}B, which has been applied to our experimental data as a check of consistency. As a first approximation to the problem, one may assume a simple lumped model with 4 resistors: two corresponding to the current along the LIPSS ($R_\parallel$) and two across them ($R_\perp$). The above-mentioned experimental configurations are simulated by injecting the current either through the terminals of $R_\perp$ (as shown), corresponding to the measurement $V_{\rm \,vdp}^{\perp}$ (``$_{\rm vdp}$'' for van der Pauw), or through the terminals of $R_\parallel$ and measuring voltage opposite to them (to obtain $V_{\rm \,vdp}^{\parallel}$). A simple calculation leads to the expressions for the measured resistance in both cases (${R_{\rm \,vdp}^{\perp}\equiv V_{\rm \,vdp}^{\perp}/I, \,R_{\rm \,vdp}^{\parallel}\equiv V_{\rm \,vdp}^{\parallel}/I}$), that may be inverted to give
\begin{eqnarray}
\label{eq:resistance_circuit}
&&R_{\perp} = 2\left(R_{\rm \,vdp}^{\perp}  + \sqrt{{R_{\rm \,vdp}^{\parallel}}{R_{\rm \,vdp}^{\perp}}}\,\right)
\nonumber
\\
\nonumber
\\
&&R_{\parallel} = 2\left(R_{\rm \,vdp}^{\parallel}  + \sqrt{{R_{\rm \,vdp}^{\parallel}}{R_{\rm \,vdp}^{\perp}}}\,\right) 
\end{eqnarray}
i.e., one obtains the resistance along each of the anisotropy principal directions.

In order to test the above idea, we applied these equations to our experimental data as illustrated in Fig.~\ref{fig:figure_A1}B; the experimental curves are deconvoluted, indicating that the peak effect comes from the difference between the transitions to the normal state along the respective channels. Recall that the appearance of a maximum in the curve $R_{\rm \,vdp}^{\perp}$ relates to the beginning of dissipation along the parallel channel, which leads to a reduction of $V_{\rm \,vdp}^{\perp}$.

In passing, we note that the curves in Fig.~\ref{fig:figure_A1}A correspond to a moderate transport current, and thus, are probing the flux pinning regime. As shown in the dataset, when measurements are performed for weak currents, one finds sharp peaks very close to the transition, corresponding to the nearly isotropic behavior found for the laser-treated samples in the reversible regime.

Finally, we mention that the anisotropic resistance model was checked for consistency by using the obtained values $R_{\perp}, R_{\parallel}$ to successfully predict $R_{\rm \,vdp}^{\perp}, R_{\rm \,vdp}^{\parallel}$ for different sets of measurements as a function of applied field and temperature.
We have found a simple representation of the intrinsic field and temperature dependencies of \( R_{\perp} \) and \( R_{\parallel} \), that have been used to check for consistency of our holistic approach in combination with Eq.(\ref{eq:resistance_circuit}) inverted to obtain the experimental data. They read
\begin{align}
\label{eq:predicted_resistance}
R_{\perp}(H, T) &= R_{\perp 0} \cdot \left[1 + e^{-\alpha_{\perp} \cdot \left(H - \left(-\beta_{\perp} T +\, \gamma_{\perp}\right)\right)}\right]^{-1} \nonumber \\
R_{\parallel}(H, T) &= R_{\parallel 0} \cdot \left[1 + e^{-\alpha_{\parallel} \cdot \left(H - \left(-\beta_{\parallel} T + \, \gamma_{\parallel}\right)\right)}\right]^{-1}
\end{align}
{with  $R_{\perp 0}$ and $R_{\parallel 0}$ the values at the plateau resistance for the perpendicular and parallel components, and $\alpha, \beta, \gamma$ fit parameters for each set of curves.}

\appsection{\label{app:appB} Influence of the threshold for determining $J_{\rm c}$}
\setcounter{figure}{0} 

The experimental determination of the transport critical current density is customarily based on a threshold electric field criterion, typically $E_{\rm c}=1\,\mu{\rm V\, cm}^{-1}$, and is usually applied to the $V(I)$ curves obtained by the four-probe technique. Then, by changing the values of the applied field and temperature, one may establish the sample's $J_{\rm c}(H, T)$ response. The actual selection of $E_{\rm c}$ is, in fact, arbitrary and frequently related to the noise level of the experimental equipment. This method is generally robust for sharp transitions, with nearly vertical voltage increase beyond the so-defined critical current. 
Other criteria, such as a resistance threshold or a linear fit and extrapolation criterion, may also be employed to mitigate the transition tail effects~\cite{Ekin}. In this section, we go into detail about the method used for obtaining the $J_{\rm c}(H,T)$ dependence based on four-probe resistive measurements. By using a threshold criterion for the recorded resistance, we pinpoint the applied current as critical for the corresponding values of $H$ and $T$. 
Specifically, a threshold value $R_{th} =\, 0.01\,R_n$ has been used, with $R_n$ the well-defined normal state resistance plateau above the superconducting-to-normal transition.
Provided that the value $R_n$ happens to be independent of the applied transport current, the equivalence of the normalized resistance threshold for the measurements on a given sample is granted. On the other hand, although $R_n$ yields different values for different samples, using relative quantities gives a measure of the physical mechanism that breaks perfect superconductivity down at a given field and temperature. For instance, it may represent the contribution of the flux flow resistance~\cite{bardeen_65}.

\begin{figure*}[!]
    \includegraphics[width=.75\textwidth]{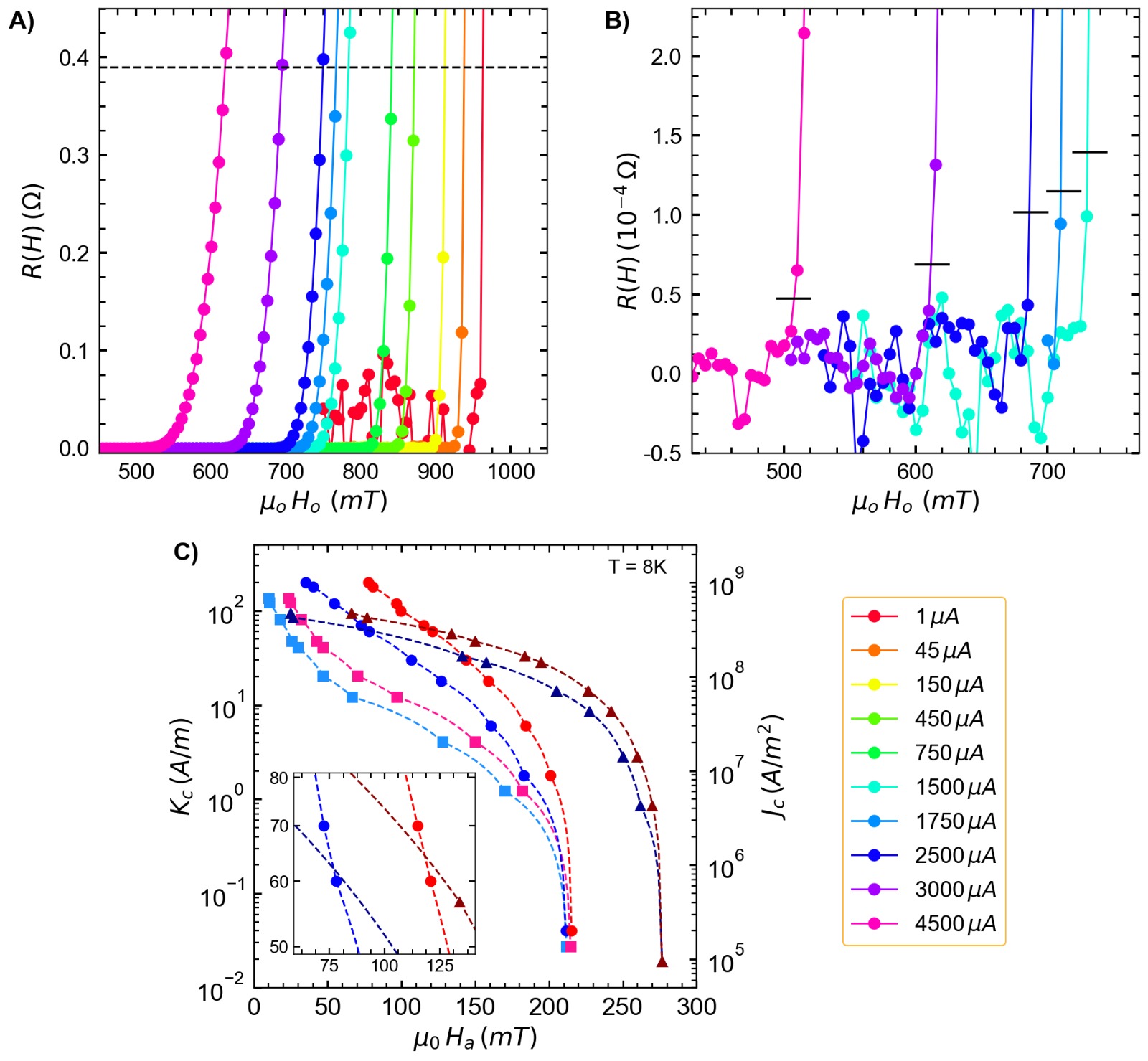}
    \caption{ A): $R-H$ curves of the BS$_{\rm PERP}$ sample for different applied currents at 6K. The dashed line represents the $0.01\,R_N$ criterion. B): Zoom of the high current R-H curves. The horizontal dashes represent the $1\,\mu{\rm V/cm}$ criterion. C): Main panel: $K_{\rm c}(H)$ at $8$~K obtained using both criteria: red colors correspond to the proportional normal resistance criterion and blue colors correspond to the microvolt criterion. Inset: Zoom into the crossover.}
    \label{fig:voltage criterion}
\end{figure*}

This criterion has been used by other groups to determine different parameters, 
such as the {\itshape zero-resistance transition} temperature $T_{\rm c}$ or upper critical fields $H_{c2}$~\cite{Fang_2017,Xiang_2021}. 

Fig.~\ref{fig:voltage criterion} displays the application of the \unit{\micro\volt}/cm and $R_{th}$ criteria to a set of $R(H, T_{_0})$ curves for different values of the transport current. 
As shown in Fig.~\ref{fig:voltage criterion}B, for the range of higher currents, the microvolt criterion lies well above the noise level and corresponds to resistance values at least three orders of magnitude below the $R_{th}$ indicator, detecting the onset of dissipation earlier. 
Eventually, this may imply differences in the derived critical parameters as shown in Fig.~\ref{fig:voltage criterion}. Thus, for the transport current $I=4500$~\unit{\micro\ampere}, one determines critical field values of $507$~\unit{\milli\tesla} or $618$~\unit{\milli\tesla} respectively.

Notwithstanding the above arguments and the possible pros and cons of each criterion, we show that our conclusions in this work barely depend on the actual choice. In fact, according to Fig.~\ref{fig:voltage criterion}C, the results are nearly criterion-independent for the region of small currents, i.e., giving a very robust estimate of the reversible properties ($H_{\rm c2}$). 

On the other hand, one may appreciate the effect of the criterion on the determination of the crossover point between the critical current curves for the samples BS0 and BS$_{\rm PAR}$. Based on a linear approximation, the crossover for both criteria occurs at $K_{\rm c,cross}\approx\, 60\,{\rm A/m}$. Nevertheless, the crossover differs in terms of magnetic fields, $H_{\rm cross} \approx\,118\, {\rm mT}$ and $H_{\rm cross} \approx\,77\, {\rm mT}$ for the $0.01\,R_N$ and $1\,\mu{\rm V/cm}$ criteria respectively, with $\Delta H \approx\, 41\,{\rm mT}$. Thus, it is clear that the choice of criterion does not modify either the appearance of a crossover nor the value of $K_{\rm c,cross}$, but it may modify the value derived for the magnetic field where the crossover takes place.
In summary, when the $R(H)$ or $R(T)$ curves are used to determine the sample's $J_{\rm c}(H,T)$ behavior from transport measurements, the selection of the threshold for characterizing the onset of dissipation may result in moderate quantitative differences. Still, it is not critical for a rational correlation between the laser processing conditions and the superconducting performance of the samples.  

\appsection{\label{app:appC}TDGL modelling}
\setcounter{figure}{0} 

\begin{figure}[t]
\includegraphics[width=0.45\textwidth]{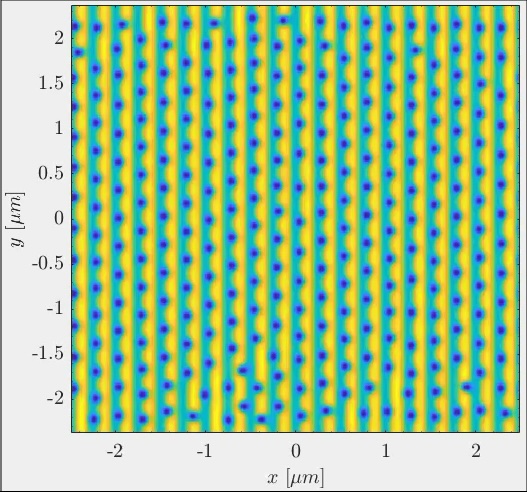}
\caption{Simulation of the current-induced flux-flow in an ideal undulated Nb thin film (${5~\mu{\rm m}\times5~\mu{\rm m}}$) with applied magnetic field along the $z-$axis (${\mu_{_0} H = 25~{\rm mT}}$) at ${T = 7.35~{\rm K}}$. Undulations run along the $y-$axis, which coincides with the direction of the external applied current. See Supplemental Material at~\href{https://drive.google.com/file/d/1yo_da32vReypFCeE64lE9ugXFhddSCp0/view?usp=sharing}{
\textcolor{blue}{this link}} for an animated video of the dynamics of this configuration.
\label{figure_S5}
}
\end{figure}

To help understand the predicted evolution of the critical current density on the applied magnetic field (presented in Fig. 9B of the main text), in this section, we provide flux flow animations for two different field values. For a better visualization of the effect of undulations on vortex motion, we present the color plot of $|\psi|^4d$ ($d$ is the film's thickness and $\psi$ the order parameter), which can be interpreted as the condensation energy density of the system. In the animations, maximum (minimum) values of this product are represented as yellow (blue). Notice that each blue circle appearing in the colour plots corresponds to a vortex, and that green bands may be identified with the valleys of the undulations, while yellow lines with the crests. 
The reduction of the calculated critical current density at large fields can be understood as follows. In the first animation (Fig.~\ref{figure_S5}), we show the vortex dynamics for an applied field of $\mu_{_0}H = 25$~mT. In this field, vortices organize themselves along the deeps of the undulations following an alternating pattern such that no two vortices occupy the same $y$-position in adjacent deeps, thus minimizing the interaction energy. To maintain this pattern during the initiation of flux flow, the applied current must be able to move the entire vortex structure at the same time, resulting in a higher critical current.

\begin{figure}[t]
\includegraphics[width=0.45\textwidth]{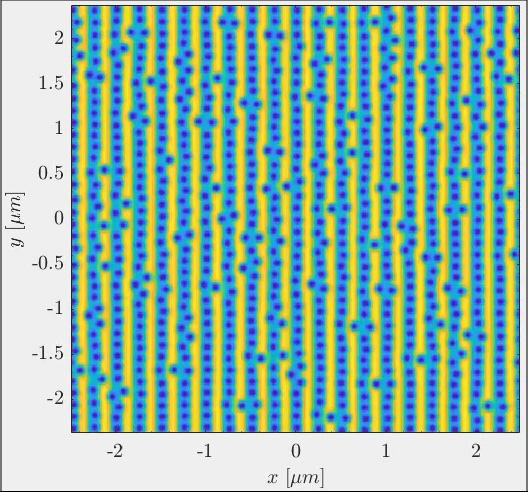}
\caption{Same as previous animation, but for $\mu_{_0}H = 65$~mT.\\
See Supplemental Material at~\href{https://drive.google.com/file/d/1p4P1AH09eRHRixt1yfvFjqbQx3NFkCZX/view?usp=sharing}{
\textcolor{blue}{this link}}.
\label{figure_S6}
}
\end{figure}

On the other hand, the second animation (Fig.~\ref{figure_S6}) shows the vortex motion for an applied field of ${\mu_{_0}H = 65~\text{mT}}$. At this higher field, the previously described vortex structure is lost, and the vortices become densely packed along the undulations. As shown in the animation, this configuration enables individual vortex hopping between adjacent deeps, thereby significantly reducing the current required to initiate flux flow.



%
%

\end{document}